\documentclass[twocolumn]{aastex63}

\usepackage{amsmath}

\shorttitle{Low states in FO Aquarii}
\shortauthors{Littlefield et al.}

\newcommand{\orbit}{$\Omega$}
\newcommand{\spin}{$\omega$}
\newcommand{\beat}{$\omega - \Omega$}
\newcommand{\dblbeat}{$2(\omega - \Omega)$}

\begin{document}

\title{The Rise and Fall of the King:\\The Correlation between FO Aquarii's Low States and the White Dwarf's Spindown }

\author[0000-0001-7746-5795]{Colin Littlefield}
\affiliation{Department of Physics, University of Notre Dame, Notre Dame, IN 46556, USA}
\author[0000-0003-4069-2817]{Peter Garnavich}
\affiliation{Department of Physics, University of Notre Dame, Notre Dame, IN 46556, USA}

\author[0000-0001-6894-6044]{Mark R. Kennedy}
\affiliation{Jodrell Bank Centre for Astrophysics, School of Physics and Astronomy, The University of Manchester, Manchester M13 9P, UK}

\author{Joseph Patterson}
\affiliation{Department of Astronomy, Columbia University, 550 West 120th Street, New York, NY 10027, USA}

\author{Jonathan Kemp}
\affiliation{Mittelman Observatory, Middlebury College, Middlebury, VT 05753, USA}
\affiliation{Visiting Astronomer, Cerro Tololo Inter-American Observatory}

\author[0000-0002-7181-2554]{Robert A. Stiller}
\affiliation{Department of Physics, University of Notre Dame, Notre Dame, IN 46556, USA}

\author{Franz-Josef Hambsch}
\affiliation{AAVSO observer}
\affiliation{Vereniging Voor Sterrenkunde (VVS), Brugge, Belgium}
\affiliation{Bundesdeutsche Arbeitsgemeinschaft f\"{u}r Ver\"{a}nderliche Sterne e.V. (BAV), Berlin, Germany}
\affiliation{CBA Mol, Belgium}

\author{Te\'{o}filo Arranz Heras}
\affiliation{AAVSO observer}
\affiliation{Observatorio Las Pegueras de Navas de Oro (Segovia), Spain}

\author[0000-0002-9810-0506]{Gordon Myers}
\affiliation{AAVSO observer}
\affiliation{Center for Backyard Astrophysics, San Mateo}
\affiliation{5 Inverness Way, Hillsborough, CA, USA}

\author[0000-0001-5888-9162]{Geoffrey Stone}
\affiliation{AAVSO observer}
\affiliation{CBA-Sierras, Auberry, CA}

\author{George Sj\"oberg}
\affiliation{AAVSO observer}
\affiliation{The George-Elma Observatory, Mayhill New Mexico, USA}

\author{Shawn Dvorak}
\affiliation{AAVSO observer}
\affiliation{Rolling Hill Observatory, Lake County, Florida, USA}

\author{Peter Nelson}
\affiliation{AAVSO observer}
\affiliation{Ellinbank Observatory, Australia}

\author{Velimir Popov}
\affiliation{AAVSO observer}
\affiliation{IRIDA Observatory, NAO Rozhen, Bulgaria}
\affiliation{Department of Physics and Astronomy, Shumen University, Bulgaria}

\author{Michel Bonnardeau}
\affiliation{MBCAA Observatory, Le Pavillon, 38930 Lalley, France}

\author{Tonny Vanmunster}
\affiliation{AAVSO observer}
\affiliation{Vereniging Voor Sterrenkunde (VVS), Brugge, Belgium}
\affiliation{CBA Belgium Observatory, Walhostraat 1A, B-3401 Landen, Belgium}
\affiliation{CBA Extremadura Observatory, 06340 Fregenal de la Sierra, Badajoz, Spain}

\author[0000-0002-1381-8843]{Enrique de Miguel}
\affiliation{Departamento de Ciencias Integradas, Facultad de Ciencias Experimentales, Universidad de Huelva, E-21070 Huelva, Spain}
\affiliation{CBA Huelva, Observatorio del CIECEM, Parque Dunar, Matalasca\~nas, E-21760 Almonte, Huelva, Spain}

\author{Kevin B. Alton}
\affiliation{AAVSO observer}
\affiliation{Desert Bloom Observatory, Benson AZ, USA}

\author{Barbara Harris}
\affiliation{AAVSO observer}
\affiliation{Bar J Observatory, New Smyrna Beach, FL}

\author{Lewis M. Cook}
\affiliation{AAVSO observer}
\affiliation{CBA Concord, Concord, CA, USA}

\author{Keith A. Graham}
\affiliation{AAVSO observer}
\affiliation{Observatory in Manhattan, IL}

\author{Stephen M. Brincat}
\affiliation{AAVSO observer}
\affiliation{Flarestar Observatory, San Gwann SGN 3160, Malta}

\author{David J. Lane}
\affiliation{AAVSO observer}
\affiliation{Astronomy and Physics, Saint Mary's University, 923 Robie Street, Halifax, NS B3H 3C3 Canada}

\author{James Foster}
\affiliation{AAVSO observer}

\author{Roger Pickard}
\affiliation{British Astronomical Association Variable Star Section}

\author{Richard Sabo}
\affiliation{AAVSO observer}
\affiliation{Pine Butte Observatory, Bozeman, Montana}

\author{Brad Vietje}
\affiliation{AAVSO observer}
\affiliation{Northeast Kingdom Astronomy Foundation, Peacham, VT, USA}

\author{Damien Lemay}
\affiliation{AAVSO observer}
\affiliation{CBA Quebec, Canada}

\author{John Briol}
\affiliation{AAVSO observer}
\affiliation{Spirit Marsh Observatory}

\author{Nathan Krumm}
\affiliation{AAVSO observer}

\author{Michelle Dadighat}
\affiliation{AAVSO observer}
\affiliation{Moka Observatory}

\author{William Goff}
\affiliation{AAVSO observer}

\author{Rob Solomon}
\affiliation{AAVSO observer}

\author{Stefano Padovan}
\affiliation{AAVSO observer}

\author{Greg Bolt}
\affiliation{AAVSO observer}
\affiliation{CBA Perth, 295 Camberwarra Drive, Craigie, Western Australia 6025, Australia}

\author{Emmanuel Kardasis}
\affiliation{AAVSO observer}
\affiliation{Hellenic Amateur Astronomy Association}

\author{Andr\'e Deback\`ere}
\affiliation{AAVSO observer}
\affiliation{3 FT user (LCO robotic telescope networks)}

\author{Jeff Thrush}
\affiliation{AAVSO observer}
\affiliation{Thrush Observatory, Manchester, MI, USA}

\author{William Stein}
\affiliation{AAVSO observer}
\affiliation{CBA Las Cruces, Las Cruces, NM, USA}

\author{Bradley Walter}
\affiliation{AAVSO observer}
\affiliation{Central Texas Astronomical Society}

\author{Daniel Coulter}
\affiliation{Department of Physics and Astronomy, Michigan State University, East Lansing, MI, USA}

\author{Valery Tsehmeystrenko}
\affiliation{AAVSO observer}
\affiliation{Heavenly Owl Observatory, Odessa, Ukraine}

\author{Jean-Fran\c cois Gout}
\affiliation{AAVSO observer}

\author[0000-0003-0828-6368]{Pablo Lewin}
\affiliation{AAVSO observer}
\affiliation{The Maury Lewin Astronomical Observatory, Glendora, CA, USA}

\author{Charles Galdies}
\affiliation{AAVSO observer}
\affiliation{Znith Observatory, Armonie, E. Bradford Street, Naxxar NXR 2217, Malta}

\author{David Cejudo Fernandez}
\affiliation{AAVSO observer}
\affiliation{El Gallinero (El Berrueco, Spain)}

\author{Gary Walker}
\affiliation{AAVSO observer}
\affiliation{Maria Mitchell Observatory, Nantucket, MA 02554, USA}

\author{James Boardman Jr.}
\affiliation{AAVSO observer}
\affiliation{Krazy Kritters Observatory, 65027 Howath Rd, De Soto, WI, 54624, USA}

\author{Emil Pellett}
\affiliation{AAVSO observer}
\affiliation{James Madison Memorial High School, Madison, WI, USA}
\affiliation{Yerkes Observatory, Williams Bay, WI 53191}


\begin{abstract}

The intermediate polar FO~Aquarii (FO~Aqr) experienced its first-reported low-accretion states in 2016, 2017, and 2018, and using newly available photographic plates, we identify pre-discovery low states in 1965, 1966, and 1974. The primary focus of our analysis, however, is an extensive set of time-series photometry obtained between 2002 and 2018, with particularly intensive coverage of the 2016-2018 low states. After computing an updated spin ephemeris for the white dwarf (WD), we show that its spin period began to increase in 2014 after having spent 27 years decreasing; no other intermediate polar has experienced a sign change of its period derivative, but FO~Aqr has now done so twice. Our central finding is that the recent low states all occurred shortly after the WD began to spin down, even though no low states were reported in the preceding quarter-century, when it was spinning up. Additionally, the system's mode of accretion is extremely sensitive to the mass-transfer rate, with accretion being almost exclusively disk-fed when FO~Aqr is brighter than V$\sim$14 and substantially stream-fed when it is not. Even in the low states, a grazing eclipse remains detectable, confirming the presence of a disk-like structure (but not necessarily a Keplerian accretion disk). We relate these various observations to theoretical predictions that during the low state, the system's accretion disk dissipates into a non-Keplerian ring of diamagnetic blobs. Finally, a new \textit{XMM-Newton} observation from a high state in 2017 reveals an anomalously soft X-ray spectrum and diminished X-ray luminosity compared to pre-2016 observations.

\end{abstract}


\section{Introduction}
\label{intro}

Despite its stature as one of the most extensively studied intermediate polars (IPs), FO Aquarii (hereinafter, FO Aqr) continues to offer fresh insight into this class of object, even four decades after its discovery. IPs are a subset of the cataclysmic variable stars (CVs), which are semi-detached binaries with an accreting white dwarf (WD) and a low-mass secondary star. The characteristic that distinguishes IPs from other CVs is the WD's magnetic-field strength, which is high enough to disrupt the accretion flow but low enough that it cannot synchronize the WD rotational period with the binary orbital period \citep[for a review, see][]{P94}.

\begin{figure*}
	\includegraphics[width=\textwidth]{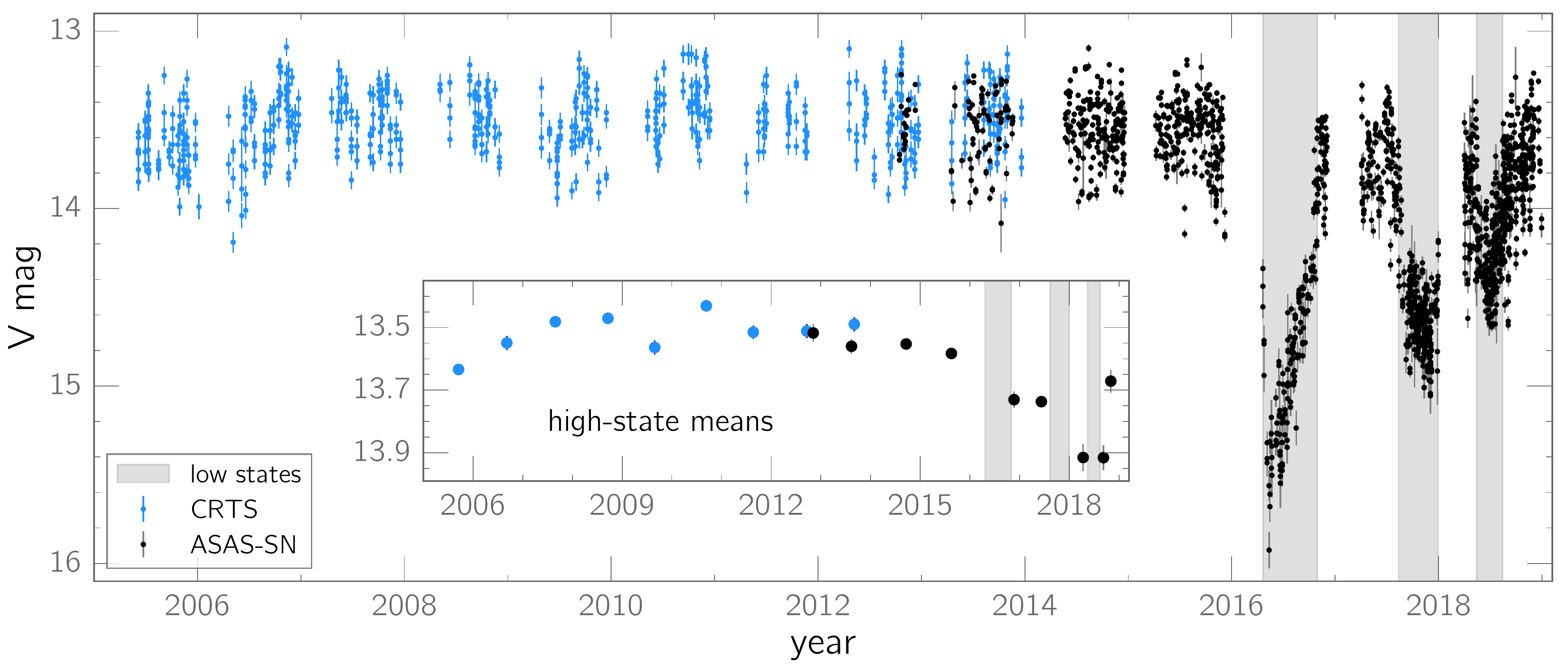}
	\caption{Long-term light curve of FO Aqr derived from observations by CRTS \citep{drake} and ASAS-SN \citep{shappee, kochanek}.The scatter is due to FO Aqr's large-amplitude, short-period variability. The inset panel shows the system's mean yearly magnitude outside of the three low states. In 2018, three means are plotted in order to better represent the pre- and post fade behavior.	In both panels, CRTS observations are unfiltered with a $V$ zeropoint, while ASAS-SN observations are primarily $V$-band, with $g$-band observations interspersed beginning in September of 2017. By comparing the mean magnitude of the system in the three bandpasses during overlapping coverage, we find that the zeropoint offsets are very small in relation to the depth of the low states, so we do not apply a zeropoint correction. Five spurious CRTS measurements have been excised from the unbinned light curve. \label{asassn_LC}}
\end{figure*}

While non-magnetic CVs transfer mass through an accretion disk, IPs can accrete from either a disk or the accretion stream from the companion star, which loses mass due to Roche-lobe overflow. The WD's magnetospheric radius determines the mode of accretion onto the WD; a large magnetosphere will disrupt the accretion stream before it can circularize into a disk, while a small one will allow the formation of a disk whose inner region is truncated at the magnetospheric radius. The former is referred to as stream-fed or `diskless' accretion, while the latter is generally known as `disk-fed' accretion. Further complicating matters is the fact that some systems show evidence of simultaneous stream-fed and disk-fed accretion. The prevailing model for these hybrid systems is that although they possess a disk, some of the accretion stream from the donor star is able to overflow the disk and collide with the magnetosphere \citep{hellier93}, although accretion from tidally induced disk structure is another possibility \citep{murray}.

Power spectral analysis is one common method of diagnosing the mode of accretion in an IP. At optical wavelengths, disk-fed accretion results in a strong signal at the WD's rotational frequency (\spin) because the inner disk is expected to be azimuthally uniform. Stream-fed accretion gives rise to signals at the orbital frequency (\orbit), the spin-orbit beat frequency (\beat), and sometimes the second harmonic of the the beat frequency, \dblbeat\ \citep{fw99}.

Irrespective of the accretion mechanism, the WD's magnetic field captures the accretion flow when the magnetic pressure exceeds the internal ram pressure of the flow. As it travels along the magnetic field lines, the gas forms a three-dimensional accretion curtain that co-rotates with the WD and impacts in an X-ray-emitting shock just above the WD's surface. Depending on the colatitude of the magnetic axis and the orbital inclination, the WD's rotation can cause a periodic variation in the viewing aspect of the accretion curtain as well as regular disappearances of the accretion shock behind the limb of the WD. These effects generate optical and X-ray pulsations, respectively, at the WD's spin frequency. Accurate timing of the spin pulsations can be used to monitor the evolution of the WD's spin period.

A number of CVs have undergone periods of diminished mass transfer. These episodes, during which the diminished accretion rate causes the system to fade by several magnitudes, are widely attributed to the passage of starspots across the secondary star near the L1 point \citep{lp94}. Low states are especially interesting in IPs because of the possibility that the accretion disk will entirely dissipate once the mass-transfer rate has dropped below a critical threshold \citep{HL17}. However, relatively few low states have been observed in IPs, with Swift J0746.3-1608 \citep{bernardini} and DO~Dra \citep{breus, am18} being two examples.

\subsection{FO Aqr}

\begin{figure*}
    \centering
    \includegraphics[width=\textwidth]{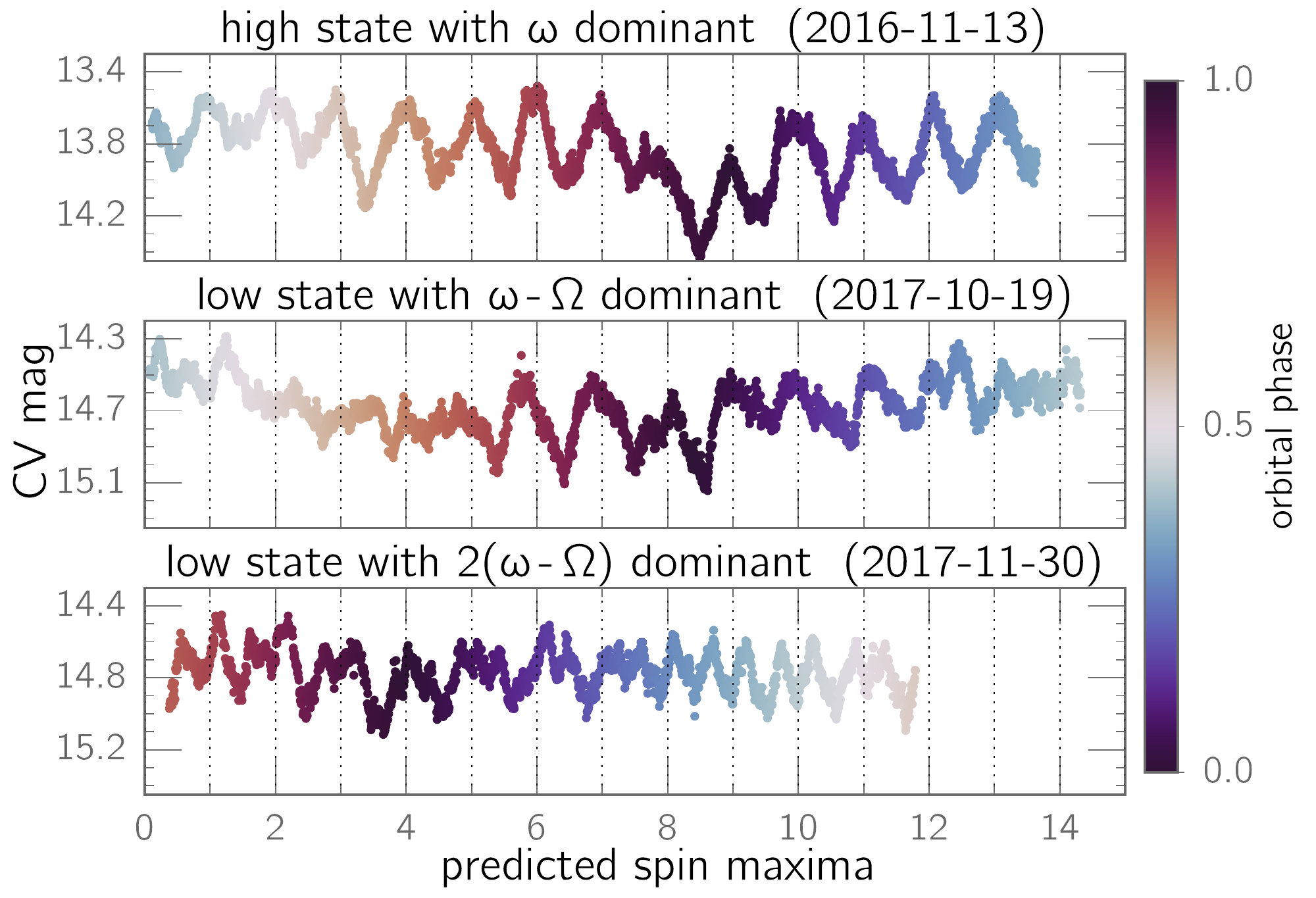}
    \caption{Representative light curves of FO Aqr in different accretion states. The dotted vertical lines indicate the expected times of spin-pulse maxima in each light curve. The vertical and horizontal scales are identical for each light curve, and the color of each point represents the orbital phase of the observation. As noted in Sec.~\ref{sec:low_states}, the relative strengths of the \spin, \beat, and \dblbeat\ frequencies change between the high and low states. The eclipses tend to be difficult to identify in any individual time series. \label{fig:sample_lightcurves}}
\end{figure*}

\subsubsection{Overview}

\citet{PS83} discovered FO Aqr as the optical counterpart of a previously known X-ray source from \citet{marshall} and concluded that it was an IP. It distinguished itself with a large-amplitude, 21-minute optical pulsation whose exceptional stability led \citet{PS83} to nickname FO Aqr ``the king of the intermediate polars.'' The binary has a 4.85-hour orbital period, punctuated by a shallow eclipse. The eclipse is visible in broadband optical photometry but not in X-rays, leading \citet{hmc89} to propose that only the outer disk is occulted. The phasing of the eclipse has remained stable for decades \citep[][and references therein]{B16}, and while the eclipse profile is normally overwhelmed by the spin pulses, phase-folded light curves of many orbital cycles suppress the spin contamination and plainly reveal the eclipse \citep[e.g.,][]{K16}. The distance to FO Aqr is $518^{+14}_{-13}$~pc, based on a probabilistic inference \citep{BJ18} from its Gaia DR2 \citep{dr2} parallax of $1.902\pm0.051$~mas.

\subsubsection{Spin period}

The spin period of the WD has received extensive attention in the literature, and the ever-increasing baseline of observations has painted a complicated picture of its evolution since the system's discovery.

It took several years after the discovery of FO Aqr to establish a sufficient baseline for \citet{pb87} and \citet{sm87} to identify a positive $\dot{P}$---\textit{i.e.}, an increasing rotational period of the WD, commonly referred to as a ``spin-down.'' \citet{om89} subsequently measured a significant $\ddot{P}$ term that indicated that the rate of the spin-down was decreasing, and \citet{sc89} found that $\dot{P} \approx 0$ in 1987. Around this time, FO Aqr entered an era of spin-up, as described by the cubic ephemerides of the spin maxima from \citet{P98} and \citet{williams}. While \citet{williams} was able to extend his ephemeris to 1998, a cycle-count ambiguity prevented him from identifying a unique ephemeris that incorporated his timing measurements from 2001 and 2002, so he proposed three possible fits to the pulse timings.

Three subsequent papers weighed in on this cycle-count ambiguity. \citet{andronov} measured the spin period in 2004 and found that it implied an extreme spin-up, far in excess of any of the three \citet{williams} fits. \citet{K16} found that the spin period during \textit{Kepler K2} observations from 2014-2015 was significantly longer than the \citet{andronov} period and interpreted this as evidence that the system had transitioned back to spin-down. \citet{B16} proposed a solution to the \citet{williams} cycle-count ambiguity and showed that the O$-$C residuals from a quadratic ephemeris implied a $\sim$25-year oscillation. However, no paper has reported an ephemeris to supersede the one from \citet{williams}.

In \citet[][hereinafter, ``Paper I'']{L16}, we muddied the waters by showing an apparent phase shift of the spin pulse in response to the system's luminosity. We speculated that the phase shift could be the result of a change in the accretion geometry but cautioned that it needed to be confirmed. Fortunately, \citet{K17} dug deeper into the issue and found that the purported phase shift was attributable to a minuscule inaccuracy in the WD spin period from \citet{K16}, propagated across two years. Although the period was constant throughout the \textit{K2} observations, there were small phase shifts of the spin pulse in response to the system's overall luminosity, and these shifts caused the Lomb-Scargle periodogram to yield a period that was inaccurate by just a few milliseconds \citep{K17}. There are two critical takeaways from \citet{K17} about the spin ephemeris: (1) the phase shift from Paper I was not of astrophysical origin and (2) more broadly, modeling a light curve of FO Aqr with a periodic function is not a reliable means of extracting either the time-averaged spin period or the time of one fiducial photometric maximum.

The latter point requires some elaboration. Many previous works have fitted a trigonometric function to an entire season of time-series photometry of FO Aqr and used this model to extract one representative time of photometric maximum, even though the underlying dataset might contain a large number of photometric maxima. However, the effect identified by \citet{K17} highlights one mechanism through which subtle variation within a dataset can thwart this type of approach. By comparison, the measurement of individual pulse timings allows a more direct and transparent means of identifying possible systematic errors, such as the correlation between pulse O$-$C and orbital phase \citep{om89}.

\subsubsection{Low states}
\label{sec:low_states}

FO Aqr is notable for undergoing deep low states caused by a diminution of the binary's mass-transfer rate. This behavior is a recent development for FO Aqr. Between 1983-2015, its overall optical luminosity remained stable, and an examination of sparsely sampled archival photographic plates obtained between 1923-1953 showed no low states \citep{gs88}; as of the end of 2015, FO Aqr had always been reported in a high state. However, when it emerged from solar conjunction in 2016 April, the system was $\sim$2 mag fainter than it had been prior to conjunction (Paper I). Additional low states occurred in 2017 \citep{L17}, 2018 \citep{L18}, and 2019 \citep{K19}. Fig.~\ref{asassn_LC} illustrates the recent spate of low states by plotting the long-term light curve of FO Aqr from the Catalina Real-Time Sky Survey \citep{drake} and the All-Sky Automated Survey for Supernovae \citep[ASAS-SN;][]{shappee, kochanek}.

Of these four low states, only the 2016 event has been examined in any meaningful detail in the literature. In Paper I, we reported optical time-series photometry of part of the low state, finding that the dominant signal in the optical power spectrum transitioned from \spin\ in the high state to \beat\ and \dblbeat\ in the low state. We interpreted this as evidence of a shift from predominantly disk-fed accretion to a stream-overflow or stream-fed geometry in the low state. We also determined that the eclipse depth decreased during the low state. However, since the recovery was still underway when the paper was published, Paper I does not describe the final stages of the low state. Lastly,
X-ray observations of the 2016 low state by \citet{K17} provided independent support for a stream-fed accretion geometry and revealed that the spectrum had softened in comparison to archival high-state X-ray observations.

\subsubsection{A note on low-state terminology}
\label{sec:definition}

There are no formal definitions for low, intermediate, and high states, so the same terminology can have subtly incongruous meanings in different papers, particularly when comparing observations obtained across the electromagnetic spectrum.

Consequently, to alleviate any ambiguity, we define a low state (or equivalently, a ``faint state'') as a period during which FO Aqr's mass-transfer rate has decreased sufficiently to change the primary mode of accretion from disk-fed to stream-fed. This occurs when the system has faded sufficiently for its optical power spectrum to show modulations at $\omega-\Omega$ and/or $2(\omega-\Omega)$ whose amplitude is comparable to that of the spin frequency $\omega$; in contrast, during the high state, $\omega$ overwhelms the other short-period frequencies. As Sec.~\ref{disk_sec} will make clear, this change in the power spectrum occurs abruptly near $V\sim14.0$, and since it is tied to a discrete physical change in the system, it provides a more objective basis for identifying low states than an arbitrary magnitude threshold. However, if power spectral information is unavailable for an observation, we define a low state as an extended period during which FO Aqr has faded below magnitude 14.0 in either the $B$ or $V$ bands.

Accordingly, the ``recovery state'' described in \citet{K17} would be called a high state after applying our criteria to the contemporaneous optical light curve, even though the X-ray luminosity during that interval remained significantly lower than usual.

\section{Data}

\subsection{APPLAUSE photographic plates}

To supplement the Harvard plates studied in \citet{gs88}, our dataset includes newly available, digitized photographic plates made available through the Archives of Photographic Plates for Astronomical Use (APPLAUSE) project\footnote{https://www.plate-archive.org/applause/}. These plates offer coverage in the interval between the end of the coverage by \citet{gs88} and FO Aqr's discovery, making it possible to explore this previously inaccessible part of FO Aqr's history. The spectral response of the plates is comparable to that of the Johnson $B$ band.

\subsection{Optical time-series photometry}

Our dataset consists of time-series photometry of FO Aqr taken during its 2016, 2017, and 2018 observing seasons, obtained with a variety of instruments. A majority of the time series were obtained by amateur astronomers in response to AAVSO Alert Notices 545, 598, and 644 \citep[][respectively]{notice1, notice2, notice3}. \footnote{ Although another low state occurred in 2019 \citep{K19}, we do not analyze it in this paper.} These data were either $V$-filtered or unfiltered with a $V$ zeropoint. All told, amateur astronomers contributed a staggering 2,870 hours of time-series photometry across the three observing seasons --- over 90\% of our time-series data.

We also obtained numerous unfiltered time series in each of the three observing seasons with the University of Notre Dame's 80-cm Sarah L. Krizmanich Telescope (SLKT) at a typical cadence of 8 sec per image.

Fig.~\ref{fig:sample_lightcurves} plots three representative time series from our dataset: one from a high state and two from low states. The AAVSO, \textit{K2}, ASAS-SN, and CRTS datasets are all freely available for download, while the SLKT and all other photometry can be obtained from either the corresponding author (C.L.) or from the journal website.

To check the internal consistency of the data, we identified overlapping light curves from different observers and checked for temporal or magnitude offsets. We identified observers who had consistent zeropoints and, as necessary, applied constant offsets to reduce zeropoint offsets. These offsets were no more than several percent.

The fundamental difficulty with measuring FO Aqr's overall brightness is the fact that its light curve is constantly fluctuating, as exemplified by the large scatter in Fig.~\ref{asassn_LC}. Because its short-period variability can exceed 0.5~mag over the course of a WD rotational cycle, a single snapshot observation of FO~Aqr can yield an inaccurate measurement of FO Aqr's overall brightness, even if the statistical error of the measurement is negligibly small. Further complicating matters is the presence of slow, apparently aperiodic variability of up to several tenths of a magnitude over the course of several days in the high state \citep{K16}. We mitigated this issue by calculating FO Aqr's average magnitude from each photometric time series.

\subsection{\textit{XMM-Newton} observations}

DDT observations of FO~Aqr were obtained using the \textit{XMM-Newton} satellite for 43~ks on 2017 May 12-13 (Obs. ID 0794580701). The EPIC-pn \citep{epic_pn} and EPIC-MOS \citep{epic_mos} cameras used a medium filter and were operated in small-window mode. The optical monitor (OM; \citealt{xmm_om}) observed FO Aqr with the UVW1 filter during the first half of the observing run and with the UVM2 filter during the second half. We analyzed the data using standard routines in the \textit{XMM-Newton} Science Analysis Software (SAS v16.1.0). All photon-arrival times were corrected to the solar system's barycenter using SAS's {\sc barycen} function.

\section{Discovery of low states in digitized photographic plates from APPLAUSE}

\begin{figure}[h]
	\includegraphics[width=\columnwidth]{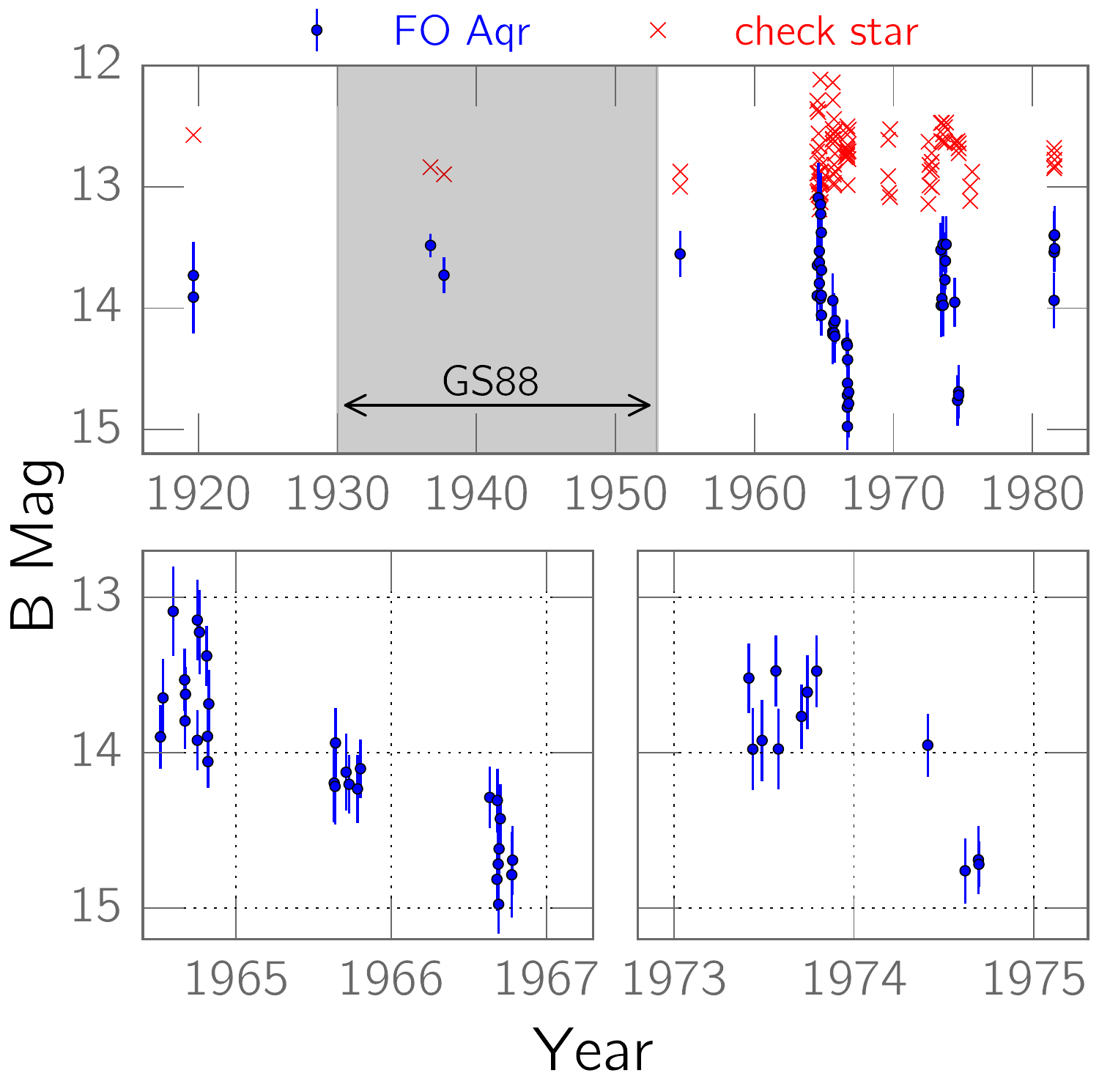}
	\caption{{\bf Top:} The APPLAUSE light curve of FO Aqr, showing low states in 1965, 1966, and 1974. The check star is UCAC4 409-138165. The shaded gray region indicates a twenty-three year span of intensive coverage in the Harvard plate archive. \citet{gs88} found no low states during that interval. {\bf Bottom left:} Closeup of the low-state observations in 1964-1965. {\bf Bottom right:} Closeup of the 1974 faint state. \label{fig:applause}}
\end{figure}

The newly available APPLAUSE observations of FO Aqr reveal previously unreported low states during 1965, 1966, and 1974, radically altering the scientific context of FO Aqr's more recent faint states. Shown in Fig.~\ref{fig:applause}, the newly identified low states had managed to evade detection until now because they occurred during the 28-year gap between the end of the \citet{gs88} light curve and the earliest reported optical photometry of FO Aqr in 1981 \citep{PS83}.

The two most conspicuous low states in the APPLAUSE light curve occurred during 1966 and 1974, when FO Aqr faded by over one magnitude to $B\sim14.7$. Moreover, in 1965, it hovered near $B\sim$14.1, approximately a half-magnitude fainter than its brightness in the high-state APPLAUSE light curve, but due to the sampling of the light curve, it is unclear whether the unusually faint measurements in 1965 and 1966 were part of the same low state. Likewise, the sampling of the low states is insufficiently dense to offer much information beyond a lower limit on their maximum depth: $\gtrsim0.5$~mag in 1965 and $\gtrsim1$~mag in 1966 and 1974.

The discovery of these low states establishes a previously unavailable historical backdrop for FO Aqr's recent low states. It refutes the assumption that FO Aqr had not experienced low states during the century leading up to 2016, a faulty inference that had been predicated upon the absence of low states in the Harvard plates from 1923-1953 \citep{gs88} and in the 34 years of observations of FO Aqr following its discovery. Instead, the 2016-19 low states are a resumption of behavior that last took place nearly a half-century ago---and not a fundamentally new stage in FO Aqr's observational history.

Could there be even more low states in unexamined photographic plates? B. Schaefer (priv. comm., 2020) searched the Harvard plate collection for observations of FO~Aqr obtained prior to 1930, the year when intensive coverage by \citet{gs88} commenced. He located and visually examined 62 plates taken between 1894 and 1923, in search of unambiguous low states. He concludes that no deep ($\gtrsim$ 1 mag) low states are apparent in the newly studied Harvard plates, thereby ruling out low states that reached or exceeded the depth of those observed in 1966, 1974, 2016, and 2017. The ongoing digitization of the Harvard plates means that a pipeline light curve of all FO~Aqr observations in that collection should be available in the reasonably near future. At that time, it will be possible to examine all of the Harvard plates in a uniform manner to search for comparatively shallow low states (similar to those observed in 1965 and 2018).

\begin{figure*}[ht]
	\includegraphics[width=\textwidth]{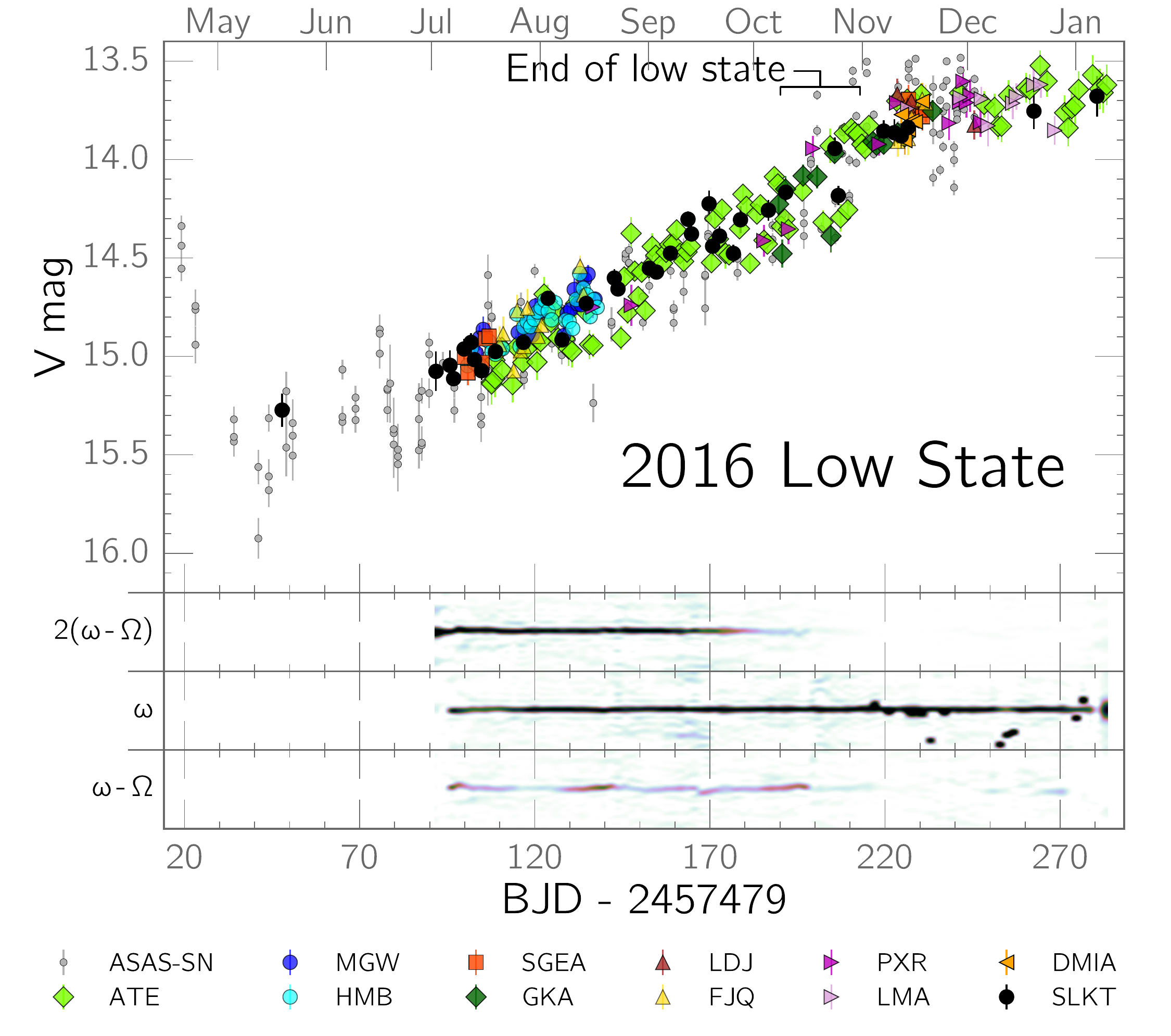}
	\caption{Light curve of the 2016 low state, with trailed power spectra of the three major short-term periodicities. Each point is the average magnitude of a time series, and each errorbar is a 1$\sigma$ estimate of the overall brightness, based on the duration of each individual time series and Monte Carlo simulations as described in the text. The different symbols correspond to codes for various observers and data sources. The beginning of each month is indicated along the top of the figure. The power at both \beat\ and \dblbeat\ drops precipitously at about the same time that the light curve jumps $\sim$0.5 mag to the high state. \label{lightcurve2016}}
\end{figure*}







\begin{figure*}[ht]
	\includegraphics[width=\textwidth]{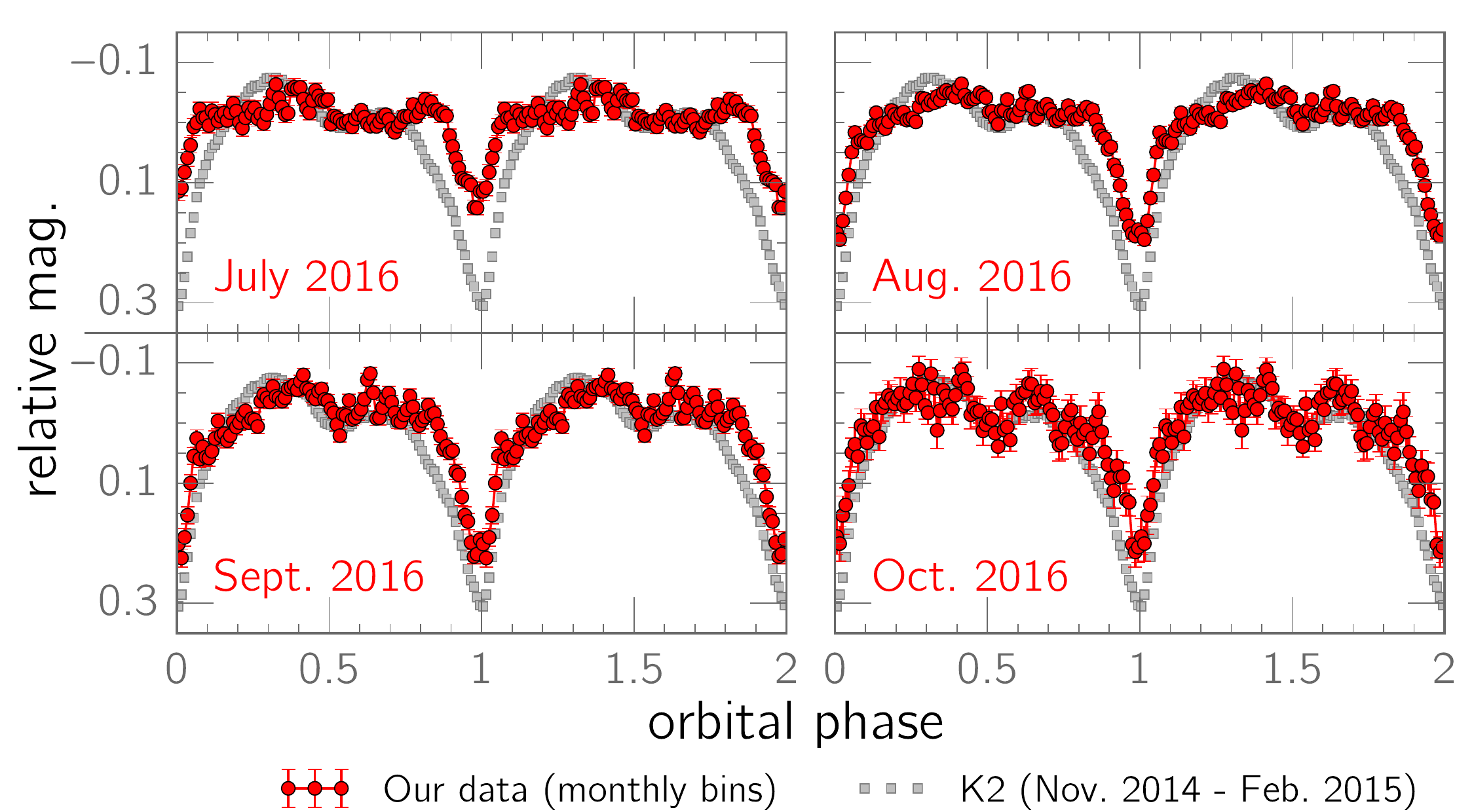}
	\caption{Phase-averaged profiles of the eclipse during the 2016 low state. The \textit{K2} orbital waveform is plotted for reference. The eclipse depth was lowest in July 2016, when the system was faintest. The presence of the eclipses throughout the low state suggests the uninterrupted presence of disk-like structure. Similar figures confirming the presence of the eclipse throughout the 2017 and 2018 low states are available from the complete figure set in the online journal. \label{2016eclipses}}
\end{figure*}

\section{Photometry of FO Aqr's different accretion states between 2016-18}

After a four-decade hiatus, FO Aqr's low states resumed in 2016, and our times-series photometry covers a total of five separate accretion states between 2016 and 2018. In this section, we describe each in chronological order. Our analysis emphasizes two aspects of the light curve: the relationship between the overall brightness and the power spectrum, and the stability of the eclipse profile throughout the low states. \citet{HL17} identified the latter as an important test of the fate of the accretion disk during a low state.

\subsection{Final recovery from the 2016 low state} \label{sec-2016}

Paper I reported time-resolved photometry through 2016 September 27, but the recovery was still underway at that time. The extended baseline of our dataset reveals when and how that low state ended.

As shown in Fig.~\ref{lightcurve2016}, the recovery entered a chaotic state in mid-September and remained that way until the end of October. During this time, the light curve showed considerable scatter, suggestive of a series of flares and dips. This behavior is present in the data of the three most prolific data sources (the SLKT and AAVSO observers GKA and ATE) during this segment of the low state. 

This behavior persisted until approximately 2016 Oct. 25, when there was a discontinuity in the light curve as the system abruptly surged to V$\sim$ 13.85. After this jump, the \beat\ and \dblbeat\ signals became very weak, with \spin\ reemerging as the dominant short-period signal. The system did not continue to brighten thereafter, so the optical recovery was complete by the end of 2016 October.

The eclipse depth gradually increased during the recovery. By phasing all data on the orbital period, we isolated the orbital modulation from the other variability in the light curve (Fig.~\ref{2016eclipses}). In July 2016, the eclipse depth was $0.16\pm0.01$ mag, and between August 1 - Sept. 30, it was $0.19\pm0.01$ mag. At no point during our observations were the eclipses absent.

The end of the 2016 low state provides a useful context for the \textit{XMM-Newton} observations of FO Aqr reported by \citet{K17}. Obtained on 2016 November 13-14, those data captured the system just over two weeks after the end of the optical low state. Although the X-ray luminosity had not returned to its pre-low-state level in those observations, the X-ray power spectrum was consistent with disk-fed accretion, as is normally seen in the high state \citep{K17}.

\subsection{The Short-Lived 2016-17 High State}

Following the completion of its recovery, FO Aqr entered into a high state at the beginning of 2016 November. ASAS-SN photometry shows that between 2016 November 1 and 2017 August 1, its average magnitude was $V = 13.72 \pm 0.02$, compared to $V = 13.565 \pm 0.008$ in three years of ASAS-SN observations prior to the 2016 low state. Each of these uncertainties is the standard error of the mean magnitude.  This 0.15-mag differential establishes that FO Aqr never fully recovered to its pre-low-state optical luminosity.

Although the 2016-17 high state was somewhat fainter than usual, it still showed the hallmarks of FO Aqr's original high state. The dominant periodicity was at \spin, and the pulsation amplitude peaked at 0.4-0.5 mag. The power spectra obtained during this period do not show any significant power at \beat, its harmonics, or other sidebands of \spin, consistent with a return to disk-fed accretion and an end to the low state.

\subsection{The 2017 Low State}

\begin{figure*}[ht]
	\includegraphics[width=\textwidth]{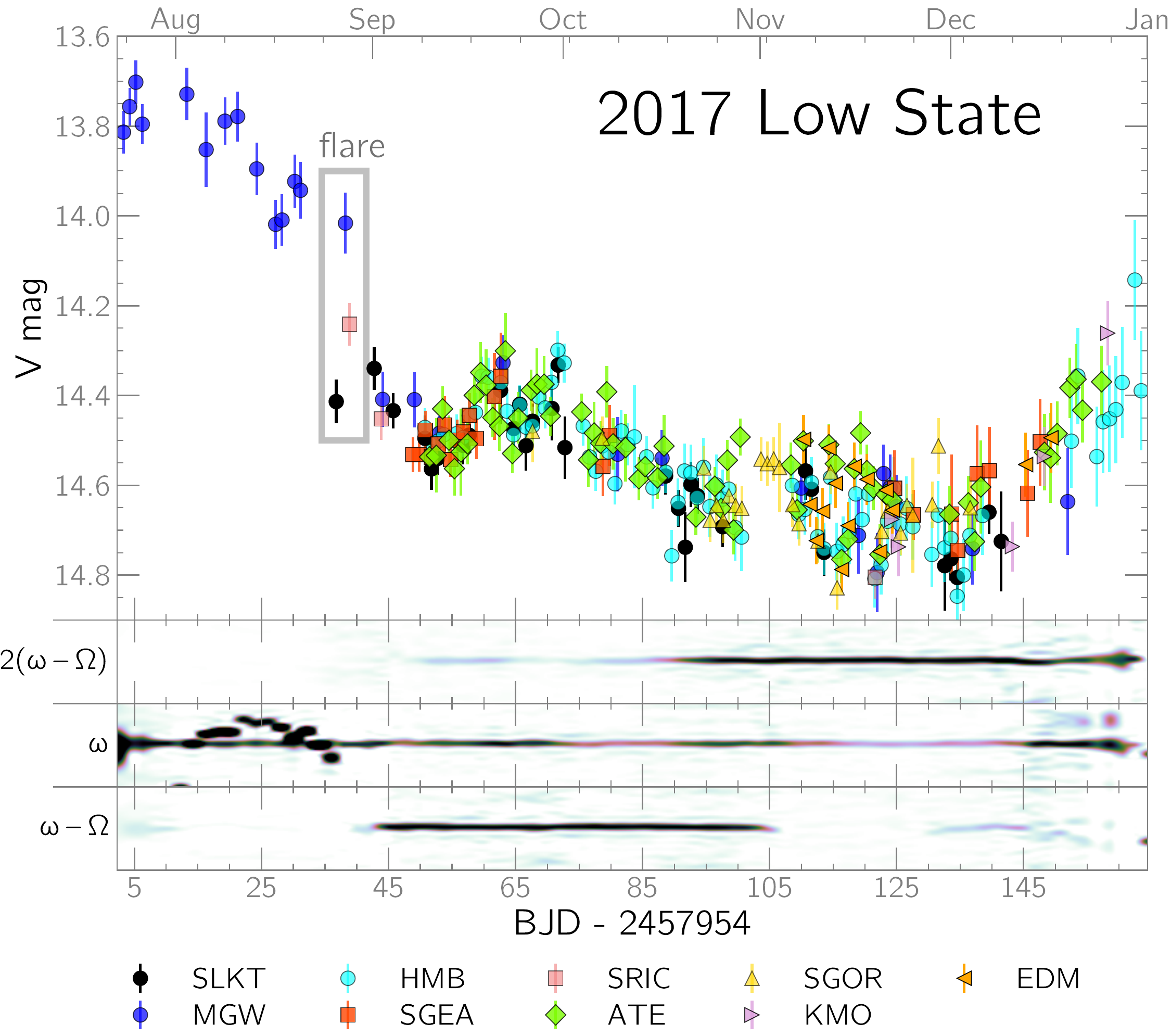}
	\caption{As with Fig.~\ref{lightcurve2016}, but for the 2017 low state. There was a gradual fade in early-to-mid August, followed by an abrupt drop into the low state. The magnitude at which this break in the light curve occurred, V$\sim$14, is very close to the magnitude at which \citet{HL17} expect the accretion disk to disappear. A brief flare occurred near the beginning of the low state, and Sec.~\ref{sec:accretion_mode} discusses how the mode of accretion changed in the $\sim$16-hour span between the MGW and SRIC observations. \label{lightcurve2017}}
\end{figure*}

Unlike the 2016 low state, which was detected only after FO Aqr had faded to its minimum brightness, the 2017 event was noticed almost immediately \citep{L17}, and there is high-quality time-series photometry of the system before, during, and after the transition into the low state. Although the coverage of the beginning was excellent, the low state was still underway when FO~Aqr reached solar conjunction, leaving the end of the 2017 low state unobserved. Fig.~\ref{lightcurve2017} shows the light curve and trailed power spectrum of the 2017 event.

Whereas the 2016 low state was characterized by a steady, gradual recovery throughout the observations, the 2017 low state featured an irregular, slow fade followed by a comparatively rapid rebrightening as solar conjunction approached. The most remarkable feature in the light curve is a $\sim$0.4-mag fade at the start of the low state whose abruptness provided a stark contrast to the laconic pace of the fading during the rest of the 2017 event.

\subsection{The 2018 Low State}

\begin{figure*}
	\includegraphics[width=\textwidth]{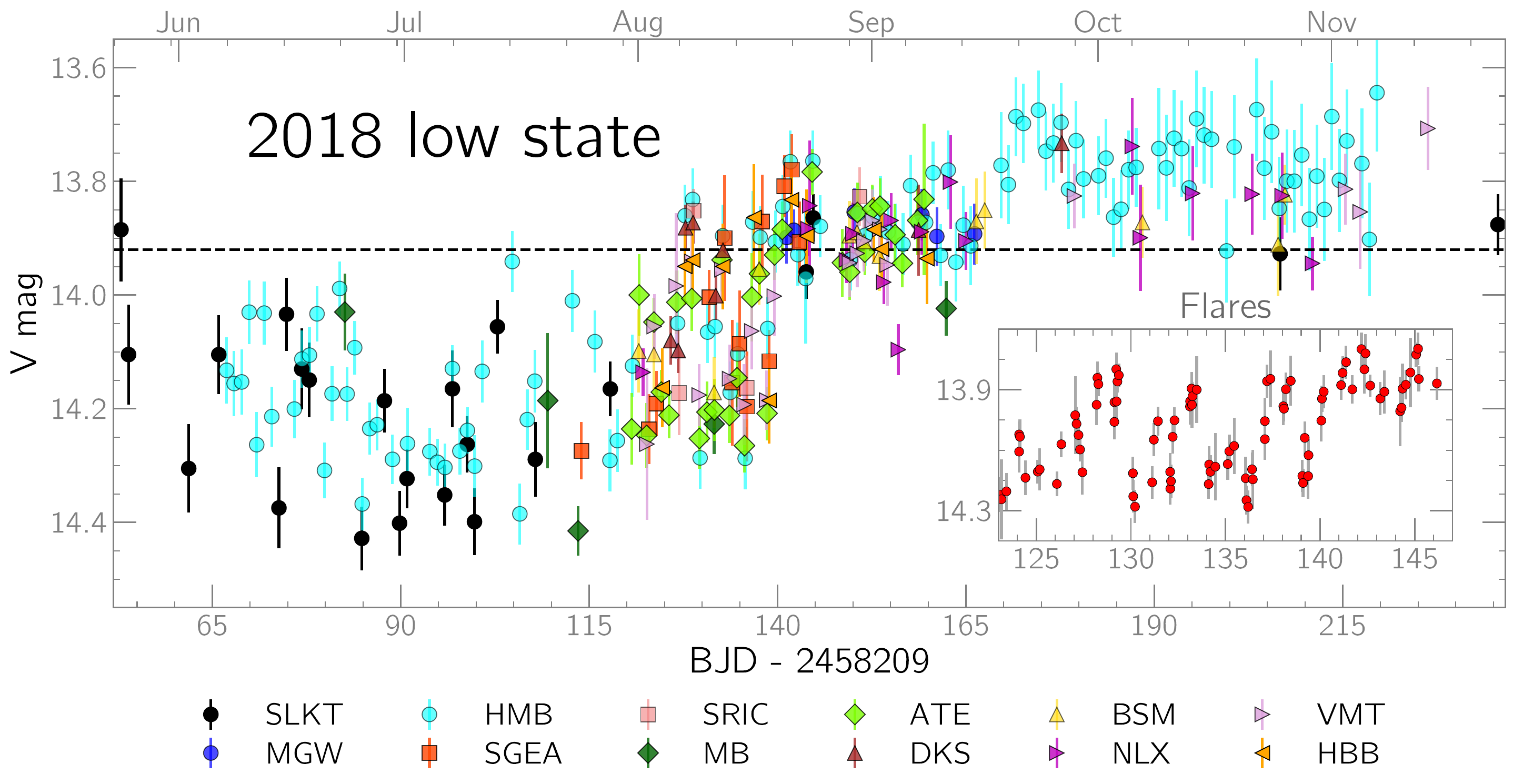}
	\caption{As with Figs.~\ref{lightcurve2016}~and~\ref{lightcurve2017}, but for the 2018 low state. The horizontal dashed line represents FO Aqr's brightness in pre-low state ASAS-SN photometry obtained in 2018. The spline fit is meant to guide the eye and models only the most densely sampled part of the light curve. We omit the trailed power spectrum because of the comparatively sporadic coverage of the 2018 low state and because the sliding window size is too wide to resolve the rapid changes observed during the $\sim$0.3-mag flares in early August, during which FO Aqr briefly reattained its pre-low-state brightness. The inset panel zooms in on these flares and plots them using just one marker style to improve their visibility. The low state ended in mid-August, and approximately one month later, FO Aqr brightened by an additional 0.2 mag. \label{lightcurve2018}}
\end{figure*}

ASAS-SN observations show that FO Aqr was in a high state as it emerged from solar conjunction in April 2018. In mid-May, it faded by $\sim$0.4 mag \citep{L18}. During this interval, ASAS-SN observed the system sporadically, but the sparse sampling precludes meaningful power spectral analysis. Time-series photometry of this low state began in earnest in late May, picked up in early-to-mid June, and became prolific at the start of August after an AAVSO campaign was launched. Fig.~\ref{lightcurve2018} presents the light curve of the 2018 low state, and it is immediately obvious that the 2018 event was briefer and shallower than its two predecessors, lasting only $\lesssim$3 months and dropping just $\sim$0.5 mag relative to its brightness before the low state.

The start of intensive AAVSO coverage in August fortuitously coincided with the system's recovery to a bright state, and the light curve showed a series of flares and dips, each lasting a few days, as the system vacillated between the low and high states. During these flares, individual time series showed an intermittent spin pulse that dominated the light curve when present. The interval between these flares was $\sim$5 days, though they were not strictly periodic.

Phase-averaged orbital light curves of the 2018 low state confirm that, as with the 2016 and 2017 low states, there was an eclipse.


\subsection{The late 2018 high state}

After the series of flares in mid-August, FO Aqr completed its recovery and remained at $V\sim13.9$ for approximately one month. During this time, \spin\ was consistently the only significant short-term periodicity. In mid-September, the system brightened by another $\sim$0.2~mag but did not show any contemporaneous changes in its power spectrum. Even after this brightening, the system was still fainter than its historical brightness of $V = 13.565 \pm 0.008$, suggesting that the accretion rate remained below what it had been in the pre-2016 high state.

\section{X-ray observations of the 2017 high state}

\begin{deluxetable}{ccc}
	\caption{Best-fit X-ray spectral parameters (2017) \label{tab:xray}}
	\tablehead{\colhead{Component} & 
		\colhead{Parameter} & 
		\colhead{Value}}

	\startdata	
	{\sc wabs} & nH & $0.13^{+0.05}_{-0.03} \times 10^{22}$ \\
	{\sc pcfabs}$_{1}$ & nH & $15.4^{+1.7}_{-1.4} \times 10^{22}$ \\
    			& cvf & $0.72^{+0.02}_{-0.04}$ \\
	{\sc pcfabs}$_{2}$ & nH & $3.4^{+0.4}_{-0.3} \times 10^{22}$ \\
				& cvf & $0.89^{+0.01}_{-0.01}$ \\
	{\sc mekal}$_{1}$  & kT & $36.6^{+10.8}_{-5.3}$ keV \\
				& norm & $0.03034^{+0.00092}_{-0.00083}$ \\
	{\sc mekal}$_{2}$  & kT & $0.118^{+0.014}_{-0.011}$ keV \\
				& norm & $0.018^{+0.013}_{-0.007}$ \\
	{\sc gaussian}     & center & $6.485^{+0.021}_{-0.024}$ keV \\
				& $\sigma$ & $0.179^{+0.040}_{-0.019}$ \\
				& norm   & 1.29$^{+0.21}_{-0.12} \times 10^{-4}$
	\enddata
	\tablecomments{The full model was {\sc wabs} * {\sc pcfabs}$_{1}$ * {\sc pcfabs}$_{2}$ * ({\sc mekal}$_{1}$ + {\sc mekal}$_{2}$ + {\sc gaussian}). In addition to the free parameters listed above, each {\sc mekal} component had four fixed parameters: nH = 1 cm$^{-3}$, abundance = 0.5, redshift = 0.0, and switch = 1.}

\end{deluxetable}

\begin{figure*}
	\includegraphics[width=\textwidth]{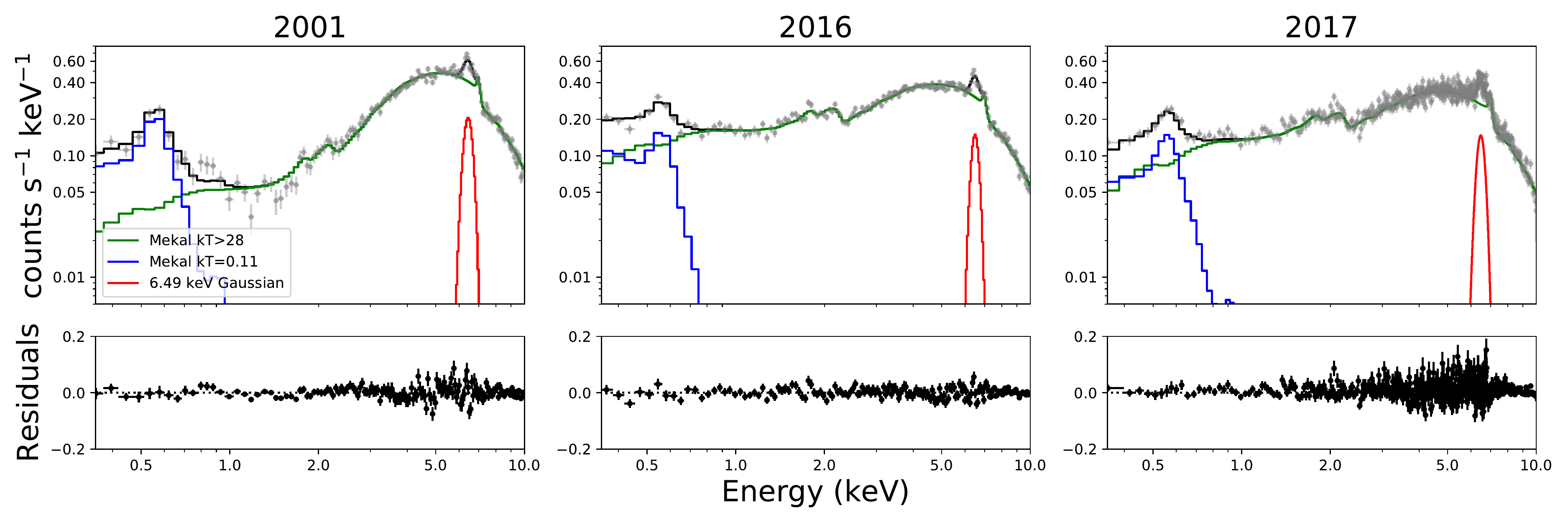}
	\caption{The high-state X-ray spectrum of FO Aqr at three different epochs, obtained with XMM Newton's EPIC-pn instrument. The spectra in the left and central panels were originally reported in \citet{evans} and \citet{K17}, respectively. The newly reported spectrum is in the right panel.\label{fig:spectra}}
\end{figure*}

Our newly reported \textit{XMM-Newton} observation of the 2017 high state bolsters our conclusion from optical photometry that FO Aqr never fully recovered to its pre-2016 state. We simultaneously fit the EPIC-pn, MOS1, and MOS2 spectra with the same model that \citet{K17} used to analyze the 2016 observation: two separate {\sc mekal} components, a {\sc gaussian} to model the 6.4 keV Fe line, an interstellar absorber, and two circumstellar absorbers, each with its own covering fraction. The 2017 spectrum is shown in Fig.~\ref{fig:spectra}, along with the 2001 and 2016 spectra for comparison. The best-fit parameters of the 2017 spectrum, which are listed in Table~\ref{tab:xray}, are statistically indistinguishable from those describing the \textit{XMM-Newton} spectrum of the 2016 high state \citep[Table~2 in][]{K17}. The similarity between the 2016 and 2017 X-ray spectra suggests that the soft X-ray excess noted in \citet{K17} might be a new, persistent feature of the X-ray light curve.

\begin{figure*}
	\includegraphics[width=\textwidth]{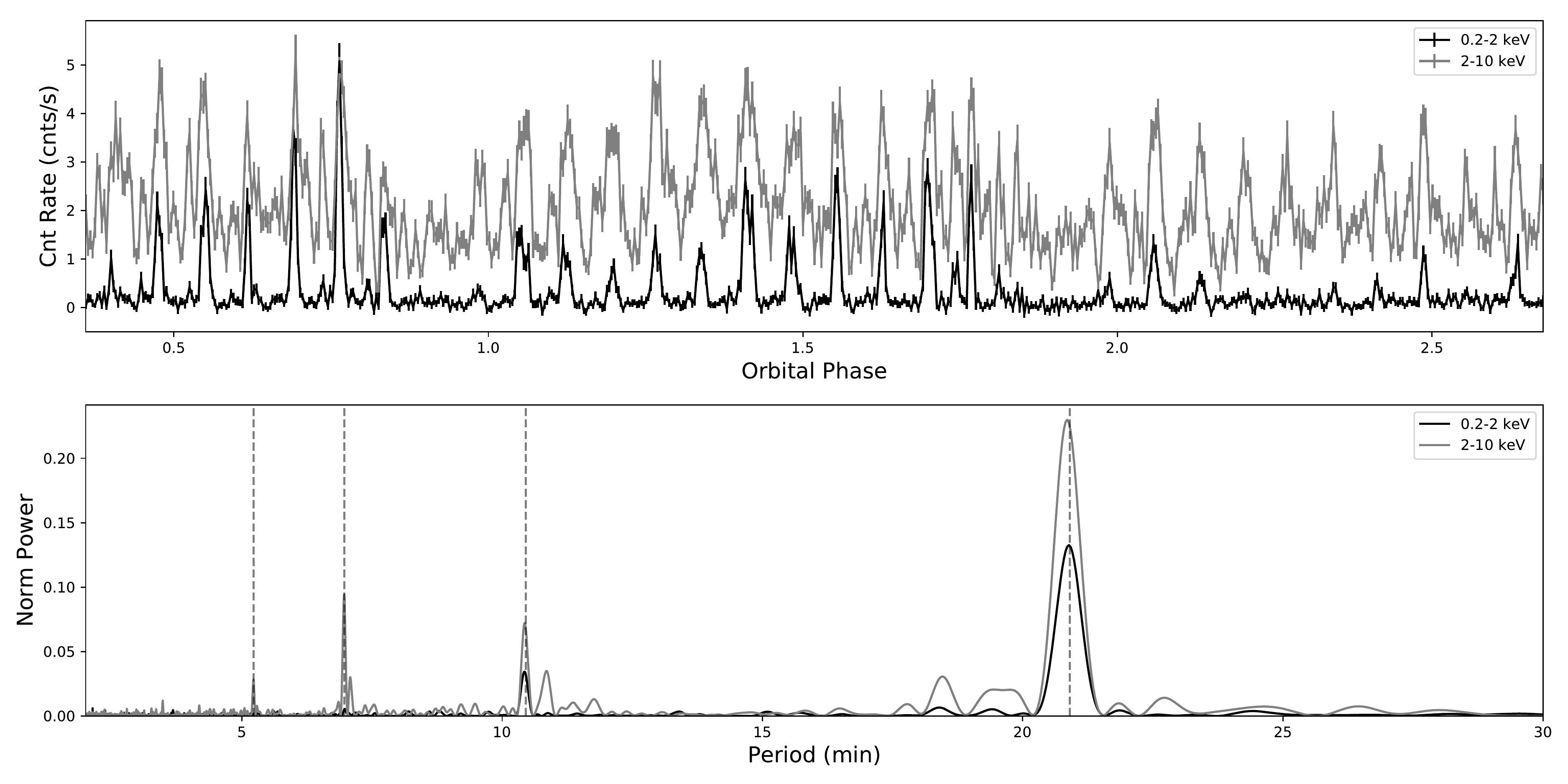}
	\caption{The 2017 XMM-Newton light curve of FO Aqr and its Lomb-Scargle power spectrum. The pulsed hard X-ray flux is significantly attenuated shortly before the predicted time of mid-eclipse. Unlike the long-term high state, soft X-ray pulses were routinely observed, albeit with an orbital-phase dependency. The strongest signal in both the hard and soft X-ray light curves is the 20.9-minute spin period. The spin signal and its first 3 harmonics are all marked with dashed vertical lines. \label{xmm_LC}}
\end{figure*}

The X-ray light curve and power spectrum (Fig.~\ref{xmm_LC}) show that the soft (0.3-2 keV) X-rays are strongly pulsed, with the count rate typically dropping to zero between spin pulses. \citet{K17} noted similar behavior, but the pulses in the soft X-ray light curve in the 2016 high state were far more sporadic than in the 2017 light curve. Indeed, during the three partial orbits covered in the 2017 light curve, the soft pulses were visible throughout the first two orbits except during orbital phases 0.8-1.0, when there was an energy-dependent drop in the pulse amplitude. This is consistent with photoelectric absorption of the soft X-rays caused by vertical disk structure, possibly related to the stream-disk collision. Since the optical eclipse is centered on orbital phase 0.0, this effect cannot be attributed to an eclipse by the secondary. A more likely candidate is a large, dense bulge in the disk. \citet{dm94} proposed such a structure in order to explain an optical \beat\ signal during a time of disk-fed accretion. The soft X-ray pulses became weaker and more intermittent during the final binary orbit in the \textit{XMM-Newton} observation, implying that the structure of the absorber can vary significantly across consecutive orbits.

These changes to FO Aqr's high-state X-ray properties provide an interesting complement to its failure to recover to its original optical brightness. Both lines of evidence lead to the conclusion that even in the most recently observed high states, FO Aqr's accretion rate has remained lower than it was prior to the 2016 low state.

\section{The evolution of the WD spin period, and its correlation with low states}

An added benefit of our photometry is that it contains many well-observed spin pulses during high states in every year since 2002. By measuring the times of the pulse maxima, we update FO Aqr's spin ephemeris and show that the spin period's time derivative ($\dot{P}$) has undergone its second observed sign change---a phenomenon that has been theoretically predicted but never observed in an IP other than FO Aqr. Equally tantalizing is the timing of this sign change, which coincided with the resumption of FO Aqr's low states.

\subsection{New spin ephemeris}

We have calculated a new rotational ephemeris that describes all spin-pulse timings since year 2002.7 without any cycle-count ambiguities. To accomplish this, we took high-state, time-series photometry of FO Aqr from the SLKT dataset, the AAVSO database, the \textit{K2} light curve, and \citet{B16}\footnote{The \citet{B16} photometry is available at \url{https://konkoly.hu/pub/ibvs/6101/6181-t2.txt}.} and extracted photometric maxima by fitting a fourth-order polynomial to the local light curve around each pulse. The fitting algorithm rejected low-quality timings, such as those caused by inadequate sampling of an individual pulse or a by smeared pulse profile that lacks a clearly defined peak. All told, this dataset contains 5,670 pulse timings, including 4,715 pulse timings from the \textit{K2} light curve, 259 from AAVSO/CBA photometry obtained in each year from 2002-2015, and 314 timings from the \citet{B16} photometry. Moreover, the combined SLKT / AAVSO dataset obtained during the bright interregnum between the 2016 and 2017 low states (Fig.~\ref{asassn_LC}) contained 353 usable pulse timings.

We used the following iterative procedure to build the ephemeris. We started by using the linear \citet{K17} spin ephemeris (which is referenced to 2014.97) to create a spin-pulse O$-$C diagram, extending it as far back as possible before baseline curvature (due to $\dot{P}$) became apparent. We then fit the pulse timings within this window with a quadratic ephemeris, plotted an updated O$-$C diagram, and extrapolated it backwards in time until curvature reappeared in the O$-$C plot. By repeatedly extrapolating a trial ephemeris towards older observations and increasing the polynomial order to flatten the O$-$C diagram, we calculated a unique ephemeris that extends back to the earliest pulse timings in 2002 of:

\begin{equation}
T_{max}[BJD] = AE^4 + BE^3 + CE^2 + P_{0}E + T_{0},
\label{eqn:ephem}
\end{equation}

where $A = (9.61\pm0.44)\times10^{-24}$, $B = (6.50\pm0.19)\times10^{-18}$, $C = (1.86\pm0.18)\times10^{-13}$, $P_0 = 0.0145177196\pm0.0000000016$~d, $T_0 = 2456977.105796\pm0.000025$ in the $BJD_{TDB}$ standard, and $E$ is the integer cycle count.

\begin{figure}
	\includegraphics[width=\columnwidth]{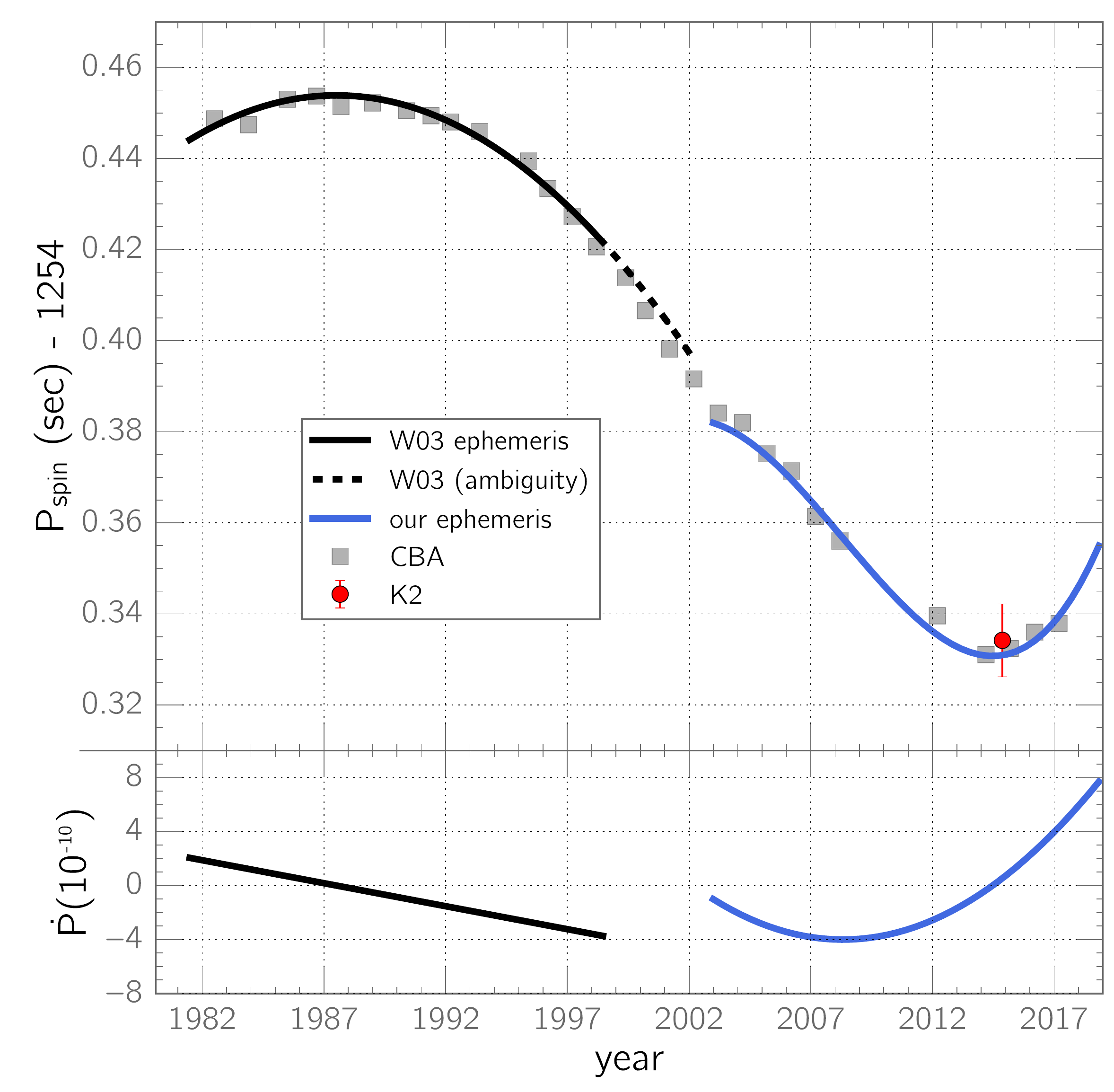}
	\caption{The spin period and $\dot{P}$ of FO Aqr since its discovery. The two solid lines are the first derivatives of different ephemerides: (1) ours, which describes spin-pulse timings obtained each year from 2002--2018, and (2) that of \citet{williams}, which describes all pulse timings from 1981--2002 (albeit with a cycle-count ambiguity after 1998). We overlay the CBA and \textit{K2} spin periods to demonstrate that our ephemeris accurately predicts them, even though they played no role in the computation of the ephemeris. \label{fig:spin_period} }
\end{figure}

Our ephemeris is free of cycle-count ambiguities. Our dataset had spin-pulse timing from each year beginning in 2002, and during our iterative procedure described in the previous paragraph, the curvature of the O$-$C never approached half of a spin cycle. Furthermore, 
Fig.~\ref{fig:spin_period} presents a comparison of the new ephemeris's predicted spin period versus the measured spin period by the Center for Backyard Astrophysics \citep[CBA; see][]{CBA} at various epochs. The agreement between the predicted and independently measured spin periods bolsters our confidence that no cycle-count ambiguity is present.

The first derivative of our spin ephemeris yields a prediction of the spin period of the WD at a given epoch, and as shown in Fig.~\ref{fig:spin_period}, it establishes unequivocally that FO~Aqr transitioned from a state of spin-up to spin-down around 2015, when the \textit{Kepler K2} observations from \citet{K16} were obtained.

Both our ephemeris and the CBA spin-period measurements strongly disagree with the \citet{andronov} value of $P_{spin} = 1254.285(16)$~s at epoch 2004.6, which they derived by fitting a sinusoidal function to time-series photometry from 10 observing runs spread across a 26-day span. As discussed previously, attempts to measure FO Aqr's spin period by representing its light curve with trigonometric functions are susceptible to systematic errors and are less reliable than O$-$C analysis \citep{K17}. We therefore believe that the spin period implied by our ephemeris and measured by the CBA more accurately describe the spin period at that epoch.

An important caveat with FO Aqr's pulse timings is that they show a great deal of unpredictable scatter, such that the residuals from any spin ephemeris are frequently $\sim\pm0.1$ in phase. Some of this is attributable to systematic contamination by the beat pulse \citep{om89}, the severity of which varies unpredictably at different epochs \citep{P98}. There is also a dependence between pulse O$-$C and the brightness of the system \citep{K16}. While a large number of pulse timings spread across many nights in a given season should cause these effects to average out, several of the observing seasons covered by our ephemeris contained only a few pulse-timing measurements that were extracted from several closely spaced time series. As a concrete example, the AAVSO dataset from 2002 contains 59 pulse timings obtained in a 7.1-day timespan. Fig.~7 in \citet{K16} shows that on such a short timescale, all observed pulses could easily experience a uniform bias of $\pm\sim$0.03 in phase, thereby mimicking the effect of an inaccurate spin period. While the resulting ephemeris might accurately predict the times of pulse maxima within the underlying dataset, it would not be measuring the actual rotation of the WD and would yield inaccurate predictions of the spin period and $\dot{P}$ near 2002.

Unfortunately, we cannot easily unify our ephemeris with that of \citet{williams}, which has a cycle-count ambiguity after 1998 due to an observational gap between 1998--2002. In principle, it is feasible to use the yearly CBA spin periods to constrain and solve this problem. However, the optimal solution is to take the CBA photometry, extract times of pulse maxima, and include them in an O$-$C analysis of all available pulse timings from the literature. Such an analysis is beyond the scope of this paper.

\begin{figure*}
	\includegraphics[width=\textwidth]{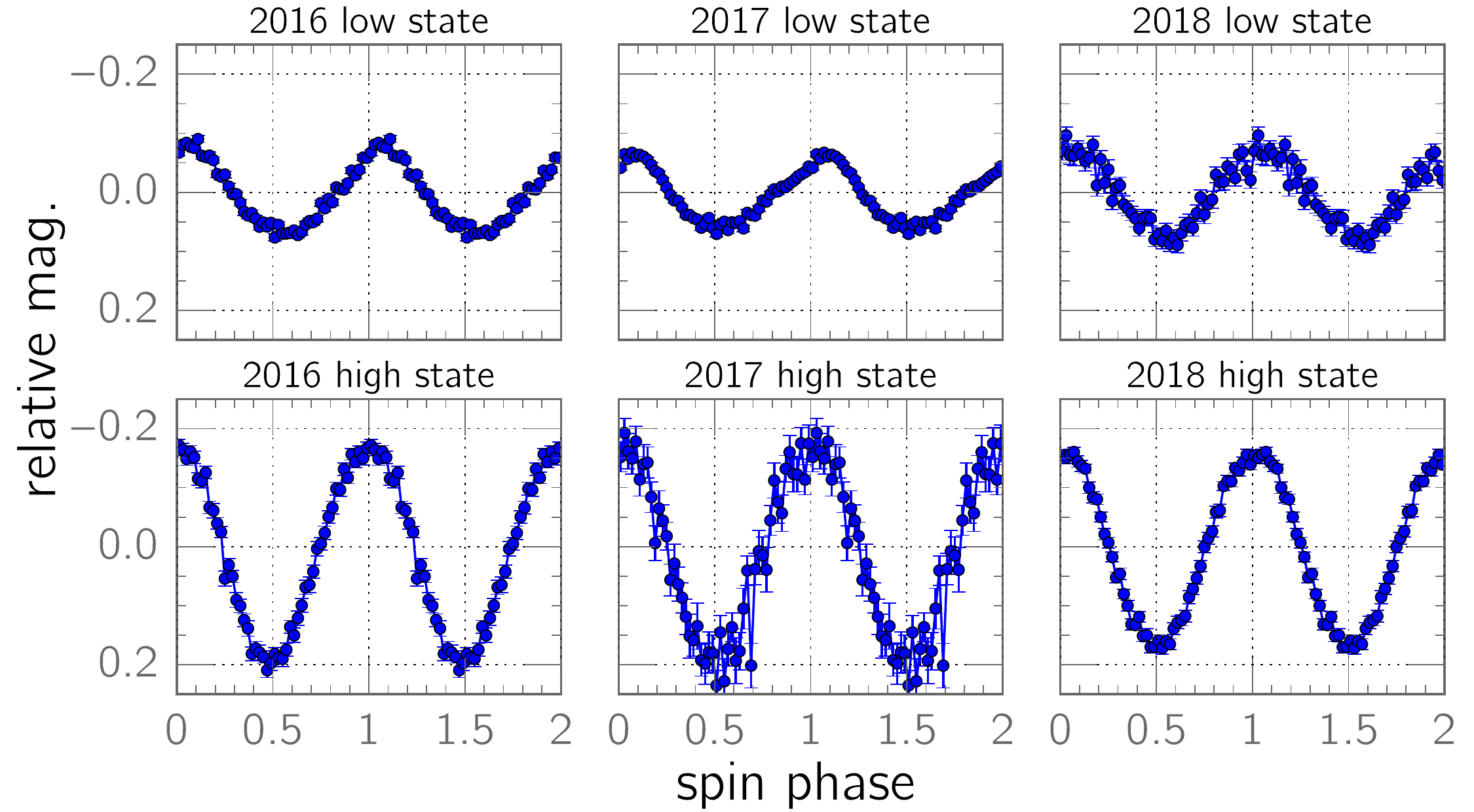}
	\caption{Phase-averaged spin profiles of each low and high state. In the 2016 and 2017 low states, the amplitude of the spin pulse was much lower, the pulse maximum was somewhat sawtooth-shaped, and the maximum occurred slightly after phase 0.0. Conversely, in each high state, the spin pulse was nearly sinusoidal and centered on phase 0.0. However, its peak-to-peak amplitude of 0.4~mag in 2016 and 2017 decreased to 0.3~mag in 2018.	\label{fig:spin_profiles}}
\end{figure*}

The spin ephemeris allows us to construct phase-averaged profiles of the spin pulse in different accretion states (Fig.~\ref{fig:spin_profiles}). We find that the amplitude of the spin pulse was lower by $\sim0.05$~mag in comparison to the 2016-17 high state. Despite the decreased pulse amplitude, the morphology of the pulse remained sinusoidal in each of the observed high states. Conversely, the spin profiles obtained during the 2016 and 2017 low states experienced a modest phase shift towards later phases ($\Delta\phi_{spin} \sim0.05$) and were asymmetric, with the rise to maximum being slower than the decline from maximum. Although previous studies \citep[e.g.][]{andronov} have noted that the amplitude of the spin pulse changes on timescales of years, the cause of these amplitude variations has never been pinpointed.

\subsection{The correlation between low states and epochs of spin-down}

\begin{figure}
    \centering
    \includegraphics[width=\columnwidth]{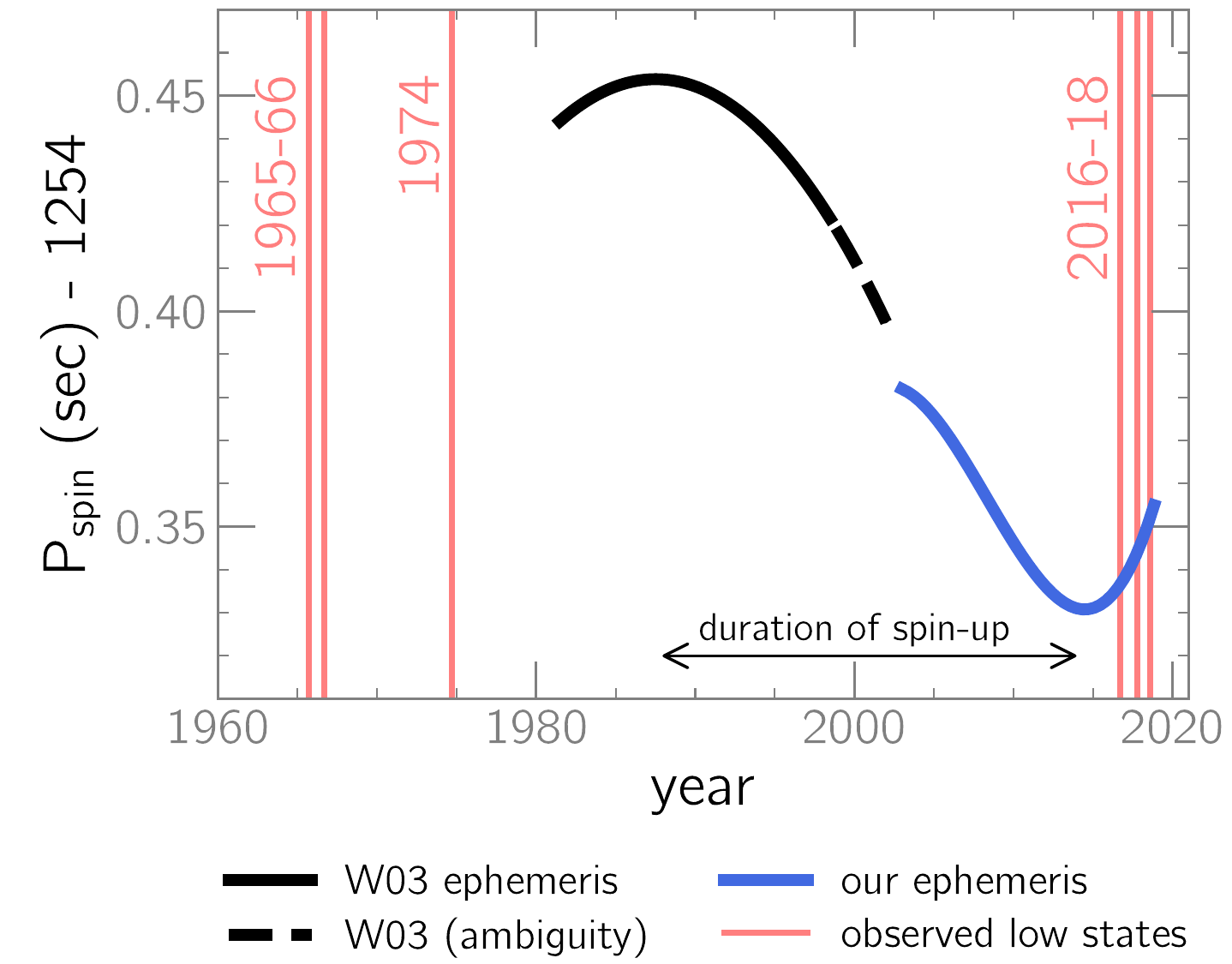}
    \caption{  The spin-period history of FO Aqr, with years of observed low states indicated. The WD was spinning down prior to 1987, and again after 2014; no low states were observed during the intervening 27-year spin-up episode. The 2016-2018 low states all occurred when the WD was spinning down. If spin-up and spin-down episodes typically last for several decades, then the newly identified low states from the 1960s and 1970s would have plausibly occurred during a state of spin-down. However, the lack of spin-pulse timings from that era means that the concomitant behavior of the spin period is unknowable.
    \label{fig:spindown_and_low_states}}
\end{figure}

The 2016-2018 low states all occurred shortly after the WD began spinning down, and no low states were observed during the 27 years of spin-up, despite good observational coverage (Fig.~\ref{fig:spindown_and_low_states}).  If we speculatively assume that spin-up and spin-down episodes in FO~Aqr last for several decades, then the 1965, 1966, and 1974 low states could plausibly have occurred during the spin-down episode that finished in 1987. We must underscore, however, that there are no spin-period measurements prior to late 1981, so it is only conjectural that the pre-discovery low states might have occurred while the WD was spinning down.

\subsection{Spin-down power}

The power generated by the spin-down of a WD is given by \citet{marsh} as
\begin{equation}
    P_{\dot{\nu}}=-4\pi^2I\nu\dot{\nu},
\end{equation}
where $I$ is the WD's moment of inertia, $\nu$ is the rotational frequency, and $\dot{\nu}$ is the rotational-frequency derivative. For an assumed WD mass of 0.8~M$_{\odot}$ and radius of 0.01~R$_{\odot}$, $I = 0.25MR^2 = 2\times 10^{43}$ kg m$^2$; from this, we calculate that $P_{\dot{\nu}} \approx\ 2\times 10^{26}\; {\rm J/s}\; \sim 0.5\; {\rm L}_\odot$ for a representative period derivative of $\dot{P} = 4\times10^{-10}$ from Fig.~\ref{fig:spin_period}.

We do not know where this energy is being deposited, but if it were driving a wind from the system, the resulting decrease in the accretion rate might help to explain why FO Aqr's low states correlate with the WD's spindown, particularly in light of calculations by \citet{HL17} that FO Aqr's high-state mass-transfer rate is only several times above the threshold at which the accretion disk would dissipate (see Sec.~\ref{disk_sec}). Any such wind would likely be detectable as a line-absorption component in ultraviolet spectroscopy.

\subsection{FO Aqr and spin equilibrium}
\label{sec:spin_discussion}

A major theoretical prediction about IPs is that angular momentum flows within the system should cause the WD's spin period to evolve to an equilibrium value, usually expressed as a fraction of the orbital period. This equilibrium is achieved when the spin-up torque from accretion equals the spin-down torque caused by the drag of the WD's magnetic field on the accretion flow, and the resulting $P_{spin} / P_{orb}$ ratio depends on whether the accretion is usually disk-fed or stream-fed \citep[e.g.,][]{kl91, ww91}. Expressed in terms of the circularization radius ($R_{circ}$), the corotation radius ($R_{co}$), and the distance from the WD to the $L1$ point ($R_{b}$), there are two types of equilibria in a diskless geometry: $R_{co} \sim R_{circ}$, which gives $P_{spin} / P_{orb} \sim 0.07$, and $R_{co} \sim R_{b}$, which yields $ 0.1 \lesssim P_{spin} / P_{orb} \lesssim 0.68,$ depending on the mass ratio \citep{kw99}. In a disk-fed geometry, the equilibrium condition is $R_{co} \sim R_{in}$, where $R_{in}$ is the inner radius of the disk \citep{kl91}. The resulting $P_{spin} / P_{orb}$ can take a wide range of values but will be smaller than the diskless $R_{co} \sim R_{circ}$ equilibrium \citep{ww91}. 

The spin period of an IP in equilibrium can exhibit small oscillations with respect to its equilibrium value if, for example, $\dot{M}$ varies coherently on timescales of years \citep{w90}. Thus, equilibrium rotation should cause $\dot{P}$ to undergo sign reversals on timescales of years as the WD alternates between episodes of spin-up and spin-down \citep{P94}. For IPs that are in spin equilibrium, the WD's magnetic moment $\mu$ can be inferred from knowledge of $P_{spin}$ and P$_{orb}$ \citep[][their Fig.~2]{N04}.

The evolution of $\dot{P}$ in FO Aqr---featuring two sign reversals only $\sim$25 years apart---offers compelling evidence that the system is in spin equilibrium. As shown in Fig.~\ref{fig:spin_period}, the spin period has exhibited a quasi-sinusoidal variation since the early 1980s, with a maximum period near 1988 and a minimum period in 2014. No other IP has experienced a sign change of $\dot{P}$, making FO Aqr an especially important system for testing theoretical predictions about the spin equilibrium phenomenon.


\section{Discussion}

\subsection{Mode of accretion}
\label{sec:accretion_mode}

\begin{figure}
    \centering
    \includegraphics[width=\columnwidth]{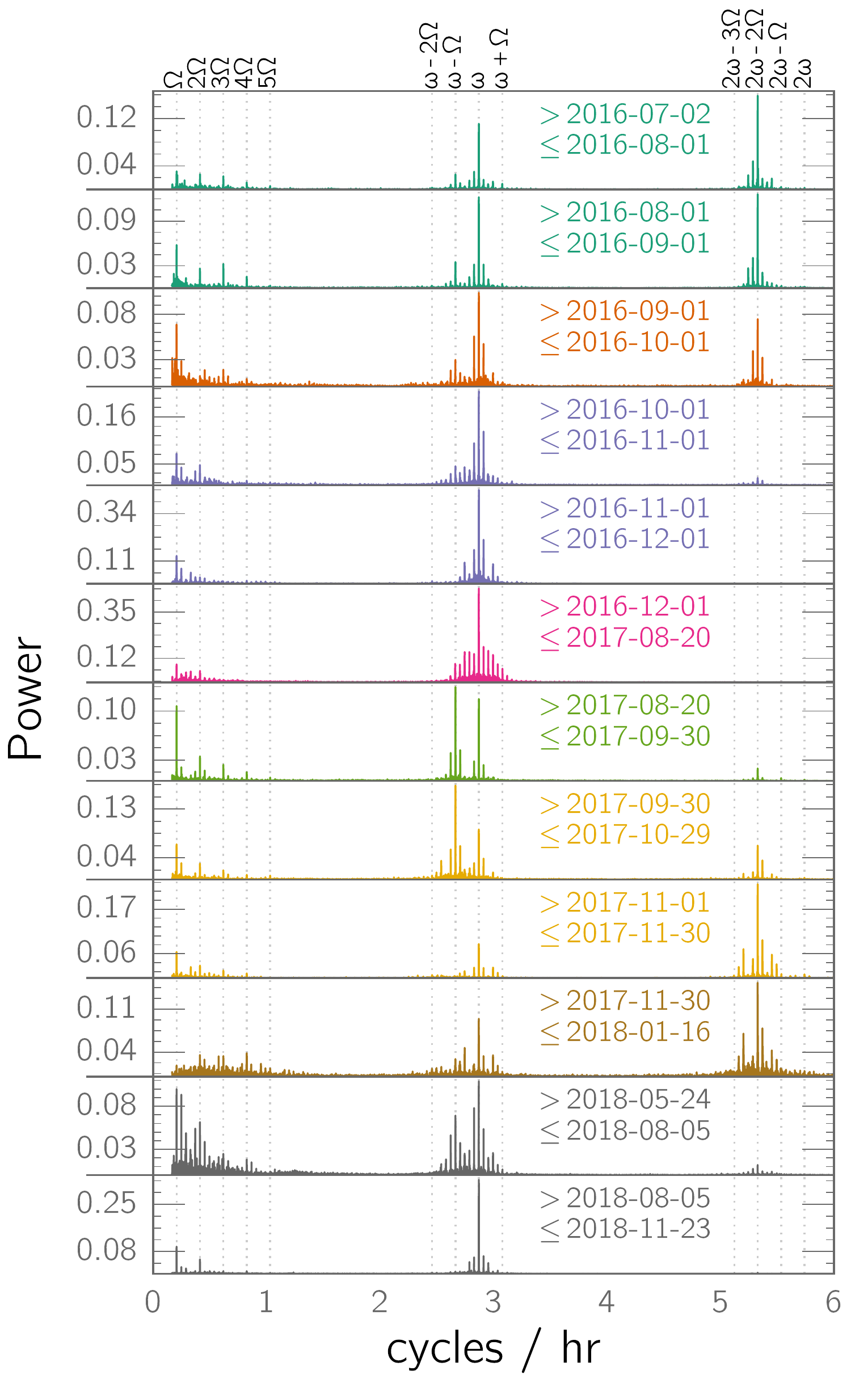}
    \caption{Lomb-Scargle power spectra of FO Aqr during different time bins. \label{fig:power_spectra}}
\end{figure}

Together, the 2016, 2017, and 2018 datasets  allow for a detailed study of the evolution of FO Aqr's power spectrum. During each transition between a high state and a low state, the system's overall brightness and its power spectrum underwent abrupt, simultaneous transitions. To underscore this point, Fig.~\ref{fig:power_spectra} plots the power spectrum from different intervals across these three observing seasons.

Observations during the deepest portion of the 2016 low state showed a very strong signal at \dblbeat\ that gradually faded away as the system rebrightened, eventually disappearing altogether when FO Aqr returned to a high state in late 2016 October. During this high state, \spin\ was the only major, short-period signal in the light curve. When the system fell into another low state in late 2017 August, \beat\ became prominent in the power spectrum, rivaling \spin. As the low state progressed, power gradually shifted from \beat\ to \dblbeat. The 2018 observations had a similar pattern; there was a significant signal at \beat\ in the low state, but it vanished during the ensuing high state, replaced by \spin.

Power spectra of individual time series suggest that the mode of accretion can change on timescales of less than one day. The flare identified in Fig.~\ref{lightcurve2017} provides one of the best examples of this behavior. Time-series photometry obtained 36 hours before the flare maximum revealed that as FO Aqr hovered near V$\sim$14.4, the amplitude of its spin pulse had decreased relative to the high state, and its power spectrum showed signals at \dblbeat\ and $\omega - 2\Omega$, neither of which had been present in the preceding high state. During the flare to V$\sim$14.0, these two new frequencies were absent, and the amplitude of the spin pulse rebounded to 0.4~mag, the value normally observed during the high state. However, just 16 hours later, the system had declined by a quarter-magnitude, and in that timespan, the amplitude of the spin pulse had plummeted, with \dblbeat\ and $\omega - 2\Omega$ both reasserting themselves in the power spectrum. As these two frequencies are associated with stream-fed accretion, the rapid changes of the power spectrum near the time of the flare suggest that the mode of accretion changed during that interval.

Another important result  of this study is that these changes in the power spectrum correlate closely with discontinuities in FO Aqr's time-averaged brightness. When \beat\ disappeared from the power spectrum in 2016, the light curve abruptly jumped by $\sim0.4$~mag, marking the end of the low state; likewise, when \beat\ reappeared in 2017, the light curve immediately dropped $\sim$0.4~mag into a low state. In both instances, there was a discontinuity in the light curve associated with a concomitant change in the mode of accretion. This suggests that the mode of accretion and the luminosity state of the system are intertwined, the implications of which we explore in Sec.~\ref{disk_sec}. 

These changes in the power spectrum are fundamentally linked to the processes by which the accretion flow couples to the WD's magnetic field. \citet{fw99} simulated the differences in optical power spectra of disk-fed and stream-fed CVs, finding that power at \beat\ and \dblbeat\ indicates that a fraction of accretion is stream-fed. Although an amplitude modulation of \spin\ at the orbital frequency can transfer power from \spin\ to \beat, this mechanism would also produce a comparable signal at the upper orbital sideband, $\omega + \Omega$ \citep{warner86}; similarly, an amplitude modulation of 2\spin\ at 2$\Omega$ would shift power to both \dblbeat\ and $2(\omega + \Omega)$. However, there is little power at the upper sidebands, indicating that neither \beat\ and \dblbeat\ can be entirely attributed to an amplitude modulation of the spin pulse and that the beat signals are instead intrinsic. 

Following this line of reasoning, Paper I determined that a significant fraction of the accretion in the 2016 low state was stream-fed, a conclusion substantiated by subsequent X-ray observations \citep{K17}. The 2017 and 2018 low states showed a similar transfer of power from \spin\ to \beat\ and \dblbeat, so we conclude that in all three low states, there was significant interaction between the accretion stream and the magnetosphere.

Although \citet{murray} proposed that tidally-induced spiral structure in the inner disk could produce optical modulations at \dblbeat\ in IPs, their mechanism would be most effective in the presence of a viscous accretion disk, a scenario that the calculations of \citet{HL17} rule out in FO Aqr's low states. Moreover, it would be difficult to explain why the spiral structure would be present in the low state but not in the high state, when the disk viscosity would be highest. Finally, the \citet{murray} model does not clearly explain why simultaneous, large-amplitude modulations at \dblbeat\ and \spin\ are never observed in FO~Aqr. We therefore prefer the \citet{HL17} disk-dissipation model (Sec.~\ref{disk_sec}) to account for the observed behavior of FO Aqr in its low states.

While the framework of \citet{fw99} offers insight into the \beat\ and \dblbeat\ signals during the low states, the nature of the signal at \spin\ during the low states presents an interesting puzzle. Paper I interpreted the presence of \spin\ during the low state as evidence that the WD accreted from an accretion disk, but subsequent calculations by \citet{HL17} show that it is unlikely that a Keplerian disk could have been present during FO Aqr's 2016 low state. As we discuss in detail in Sec.~\ref{disk_sec}, we interpret the continual presence of \spin\ as evidence that during the low states, some accretion occurred from an azimuthally symmetric, circumstellar structure (though not necessarily the Keplerian accretion disk presumed by Paper I).

\subsection{The likely dissipation of the disk}
\label{disk_sec}

\citet{HL17} predicted that the accretion disk dissipated during the 2016 low state, a scenario that could explain the simultaneous discontinuities in the optical light curve and power spectrum. Motivated by the lack of dwarf-nova outbursts during FO~Aqr's 2016 low state, they modeled the thermal state of the accretion disk during the low state and considered three possibilities: that the disk is always in a cold state, that it is always in a hot state, and that it is in a hot state but dissipates during the low state, causing the accretion stream to directly impact the magnetosphere. They showed that the first option was implausible and that although the second option was a remote possibility, it would require (1) the high-state $\dot{M}$ to exceed the expected evolutionary value by two orders of magnitude and (2) FO~Aqr's distance to exceed 1~kpc, highly inconsistent with the subsequently determined Gaia DR2 distance of $518^{+14}_{-13}$~pc. Instead, they argued that the disk radius gradually shrank towards the circularization radius until it disappeared, an event predicted to occur when the system's optical magnitude dropped to $V\sim14$ (based on the parameters assumed in their Fig.~5).\footnote{ Their assumed parameters were all independent of our Paper I.}

It is essential to note that the ubiquitous term ``diskless accretion'' is a misnomer; both observation \citep[e.g., V2400 Oph:][]{v2400oph} and theory \citep{kw99} show that nominally diskless accretion in an IP can produce a ring-like structure around the WD. Unlike a bona-fide accretion disk, however, this ring consists of diamagnetic blobs whose motion is non-Keplerian and heavily influenced by a drag force exerted by the WD's magnetic field \citep{kw99}. As a result, a ``diskless'' accretion geometry can exhibit many of the hallmarks of disk-fed accretion, such as eclipses of a disk-like structure and a spin pulse. Thus, the presence of an orbital eclipse in phase-averaged light curves of the low states (Fig.~\ref{2016eclipses}) is not prima facie evidence of an accretion disk, as the eclipses tell us nothing about the velocity field of the eclipsed matter. Because of the counter-intuitive nature of the term ``diskless accretion,'' we will hereafter refer to it as blob-fed accretion.

Returning to the data, the disk-dissipation model offers a coherent explanation for many of the key photometric properties of FO Aqr's three low states. One of our major findings is that in each well-observed transition into or out of a low state, there was a discontinuity in the light curve near $V\sim14.0$, accompanied by a simultaneous change in the power spectrum (\spin-dominated for $V\lesssim14.0$ and \beat-dominated for $V\gtrsim14.0$). \citet{HL17} predicted that at $V\sim14$, the corresponding $\dot{M}$ would be low enough to cause the magnetospheric radius to exceed the circularization radius, disrupting the disk. The exact value of the predicted magnitude at which the disk dissipates is subject to some uncertainty, owing to the theoretical ambiguities of magnetic accretion and to some of the poorly constrained physical parameters of FO~Aqr. However, the key point here is that \citet{HL17} expect the disk to dissipate at some characteristic magnitude, and the pronounced changes in FO~Aqr's photometric behavior near $V\sim14.0$ are consistent with their prediction.

In a blob-fed geometry, it is unlikely that an orbiting ring of blobs would stop the accretion stream from impacting the magnetosphere. Consequently, a blob-fed geometry would naturally account for the abrupt switch from a \spin-dominated optical power spectrum to one with greatly elevated power at \beat\ and \dblbeat. Moreover, it is natural to expect that when the accretion flow transitions from a viscous, disk-fed regime to being blob-fed, there could be a sudden drop in the optical luminosity, similar to the one that we observed. Conversely, the stream-overflow model (which presumes the presence of an accretion disk) does not offer a clear reason why there should be a break in the overall light curve or why it should coincide with a major and abrupt change in the power spectrum.

In addition, an accretion disk would respond to the diminished $\dot{M}$ on its viscous timescale \citep[$\sim$10~d per Fig.~3 in][]{HL17}, so we would expect the dissipation of the disk to be preceded by a gradual fade on a similar timescale. We observed exactly this behavior prior to the start of the 2017 low state, as FO~Aqr faded by 0.2 mag over the course of $\sim$15~d immediately before the discontinuity in its light curve (Fig.~\ref{lightcurve2017}). There was no change in the power spectrum during this interval, implying that disk-fed accretion continued unabated.

Following the end of the 2018 low state, the system showed similar behavior, except in reverse; the light curve leveled off for $\sim$four weeks before brightening by $\sim$0.2 mag (Fig.~\ref{lightcurve2018}). It is possible that once FO Aqr achieved the critical $\dot{M}$ to reestablish a viscous disk, it took longer for $\dot{M}$ to complete its recovery and replenish the disk.

It would be fruitful if a future paper were to identify concrete observational changes resulting from disk dissipation---with an emphasis on how they might be distinguished from those associated with a partially depleted accretion disk. For example, in a blob-fed geometry, we might expect to see quasi-periodic oscillations caused by beats between the WD spin and the decaying orbits of the blobs.

\subsection{Orbital waveform and eclipses}
\label{sec:eclipses}

The gradually increasing eclipse depth and width during the recovery from the 2016 low state (Fig.~\ref{2016eclipses}) are both consistent with an increase in the size of the eclipsed source. No comparable trend was present in the 2017 or 2018 low-state eclipse profiles, possibly because of the shallow depth of those low states (Fig.~\ref{asassn_LC}).

The eclipse profiles strongly disfavor the possibility that the disk experienced a thermal transition as the system migrated between the low and high states, consistent with the lack of dwarf-nova outbursts. For example, when the accretion disk in the dwarf nova U~Gem is on the cold branch, grazing eclipses of it are deep and well-defined because the stream-disk hotspot contributes a disproportionately large amount of light in comparison to the rest of the optically thin disk \citep{ugem}. However, when the disk experiences an outburst, the hotspot's relative contribution to the light curve becomes small. The eclipses shift towards earlier phases, becoming poorly defined and much shallower than in quiescence \citep{ugem}. The absence of similar behavior in FO Aqr's eclipses as the system recovered from the low state into the bright state leads us to conclude that the disk did not undergo a thermal transition during our observations.

Our referee pointed out that if the disk dissipates in the low states (as we argued in Sec.~\ref{disk_sec}), it is surprising that the low- and high-state eclipse profiles are so similar. We agree that individual eclipse light curves during the low state would probably show a large difference compared to high-state eclipses. However, the heavy contamination from the spin and beat signals makes it extremely difficult to reliably compare individual eclipses, which is why we have resorted to showing the phase-averaged eclipse profile for a given month. Although this technique  suppresses the contamination from non-orbital, periodic signals, one of its side-effects is that genuine cycle-to-cycle variations in the intrinsic eclipse profile will be averaged out. We therefore cannot reliably test whether the profiles of individual low-state eclipses differ from those of high-state eclipses.

\subsection{Flares near the beginning and end of the low states}

FO~Aqr shows several photometric flares consistent with brief spurts in the accretion rate near the beginning and end of the low states. This behavior was especially conspicuous in the 2018 light curve (Fig.~\ref{lightcurve2018}), which showed three discrete flares, each with an amplitude of $\sim$0.25~mag, when the system had recovered to V$\sim$14.2. During these flares, the light curve tended to show the strong spin pulse that traditionally accompanies its high states, but between the flares, the variability in the light curve became erratic. Another example is the aforementioned flare at the start of the 2017 low state (Fig.~\ref{lightcurve2017} and Sec.~\ref{sec:accretion_mode}).


The flares might be a manifestation of the instability that \citet{st93} proposed for accretion onto magnetized, compact stars. In their model, if the magnetosphere rotates faster than the Keplerian frequency of the inner disk, a centrifugal barrier can arise, inhibiting accretion and causing a buildup of matter just outside the magnetosphere. Eventually, the accumulated matter pushes inward and accretes abruptly. Once this reservoir of material is depleted, the cycle repeats itself. The recurrence time for these outbursts is approximately the viscous timescale at the inner rim of the disk. \citet{st93} noted that their model could also apply to IPs, and \citet{scaringi} invoked it to explain a series of recurring bursts of accretion in \textit{Kepler} observations of a low state of a weakly magnetic system, MV~Lyr.

The recurrence interval for the \citet{st93} mechanism corresponds with the viscous timescale of the region of the disk that participates in the instability. Since the instability is an inner-disk phenomenon, the expected timescale would therefore be shorter than the $\sim$2-week viscous timescale computed by \citet{HL17} for the outer disk. Thus, the $\sim$5-day interval between the flares at the end of the 2018 low state is plausibly compatible with the \citet{st93} instability.

\section{Conclusion}

Our major findings are as follows.

\begin{itemize}

	\item Digitized photographic plates from APPLAUSE reveal that FO Aqr experienced previously unknown low states in 1965, 1966, and 1974. 
	
	 \item In our time-series photometry of FO Aqr's 2016, 2017, and 2018 low states, there was a fundamental link between the accretion rate and the mode of accretion. When brighter than $V\sim14$, FO Aqr's light curve tended to be spin-dominated with only a minimal contribution from the beat frequency; when fainter, the amplitudes of the spin and beat frequencies became comparable.

    \item The eclipse was present in phase-averaged orbital profiles throughout the 2016-18 low states, meaning that a disk-like structure was present at all times during our observations. It is unclear from the eclipses whether this structure was a viscous, Keplerian accretion disk or a ring of diamagnetic blobs.

    \item FO Aqr never fully recovered to its pre-2016 optical brightness, suggesting that the mass-transfer rate remained lower than its historical level.

	\item A series of photometric flares was observed near the end of the 2018 low state. We consider the possibility that they are the consequence of the \citet{st93} instability.

	\item Our spin ephemeris links all pulse timings obtained since 2002 and establishes that the recent fusillade of low states began shortly after the WD reverted to a spin-down state in 2014.

	\item The X-ray spectrum of FO Aqr during the 2017 high state was unchanged from the 2016 high state. In both epochs, the spectrum had a significant excess of soft X-rays.

\end{itemize}

\acknowledgments

\section*{Acknowledgments}

We are grateful to Brad Schaefer for analyzing the Harvard plates of FO~Aqr on our behalf.

We thank Jean-Pierre Lasota and Jean-Marie Hameury for their comments on an earlier version of this manuscript, as well as our anonymous referee for a diligent and well-reasoned report.

This work has made use of data from the European Space Agency (ESA) mission {\it Gaia} (\url{https://www.cosmos.esa.int/gaia}), processed by the {\it Gaia} Data Processing and Analysis Consortium (DPAC,
\url{https://www.cosmos.esa.int/web/gaia/dpac/consortium}). Funding for the DPAC has been provided by national institutions, in particular the institutions participating in the {\it Gaia} Multilateral Agreement. MRK is supported by a Newton International Fellowship provided by the Royal Society. We would like to thank Norbert Schartel and the \textit{XMM-Newton} OTAC for granting us ToO observations of FO Aqr in 2017.

The Center for Backyard Astrophysics is supported in part by NSF award AST-1615456.

Funding for APPLAUSE has been provided by DFG (German Research Foundation, Grant), Leibniz Institute for Astrophysics Potsdam (AIP), Dr. Remeis Sternwarte Bamberg (University Nuernberg/Erlangen), the Hamburger Sternwarte (University of Hamburg) and Tartu Observatory. Plate material also has been made available from Th\:uringer Landessternwarte Tautenburg.

\software{Astropy \citep{astropy13, astropy18}}

\facilities{AAVSO, Gaia, Kepler, XMM}

\appendix

\section{Stroboscopic orbital light curves}

One challenge with studying FO Aqr the differential rotation of the WD makes it difficult to study the spin and orbital variations independently of one another. \citet{marshduck} employed a novel solution to this problem using many spectra across multiple orbits. By carefully binning this large dataset, they were able to study the spectrum at fixed orientations of the WD relative to the companion star --- essentially ``freezing" the WD's rotation with respect to the donor star, similar to a stroboscope. While the application of this technique to our photometric dataset does not yield any results that are central to our major conclusions, it nevertheless offers insight into the interaction between FO Aqr's spin pulse and orbital waveform. We therefore present this analysis separately from the main text.

\begin{figure}[h]
    \centering
    \plottwo{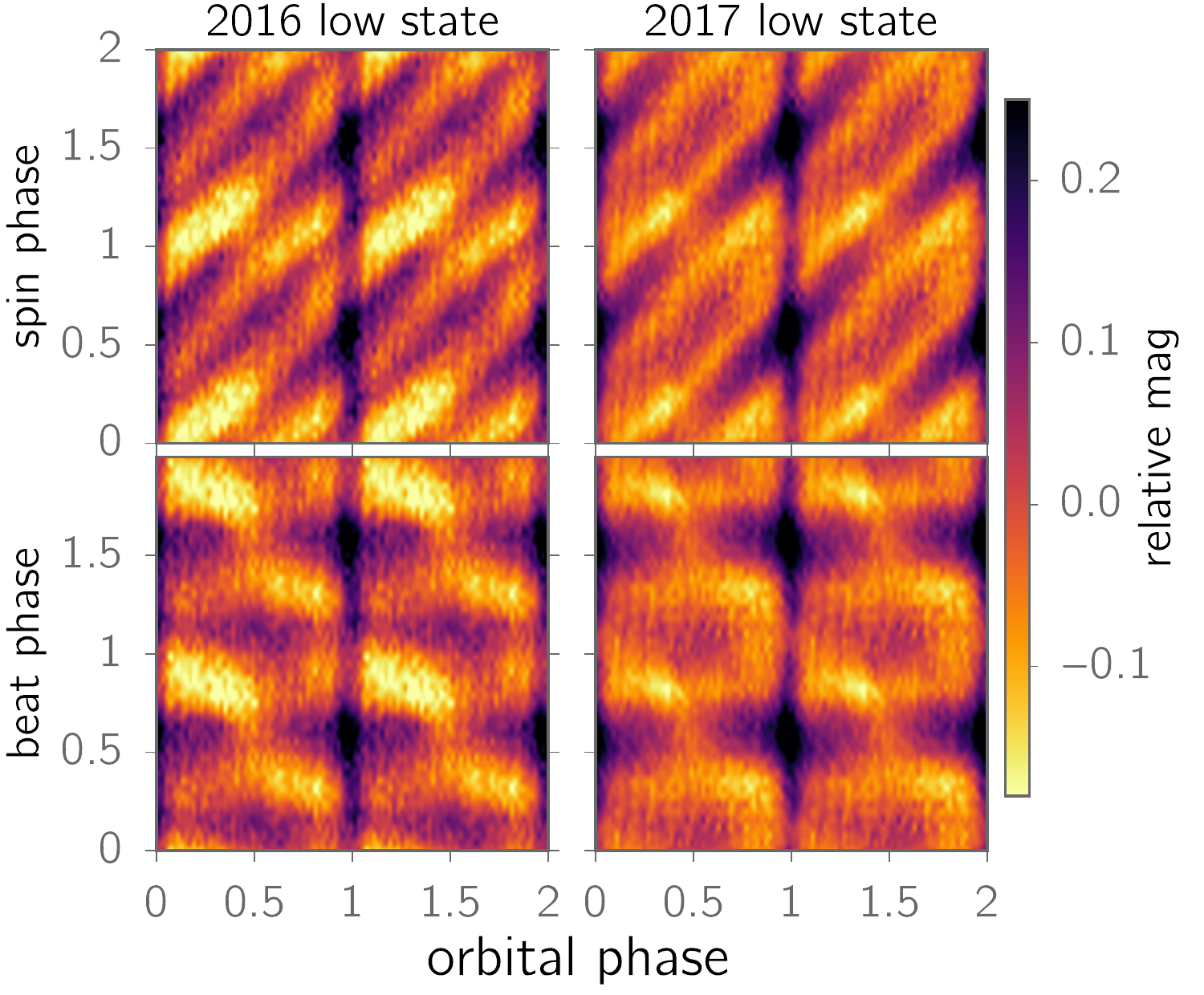}{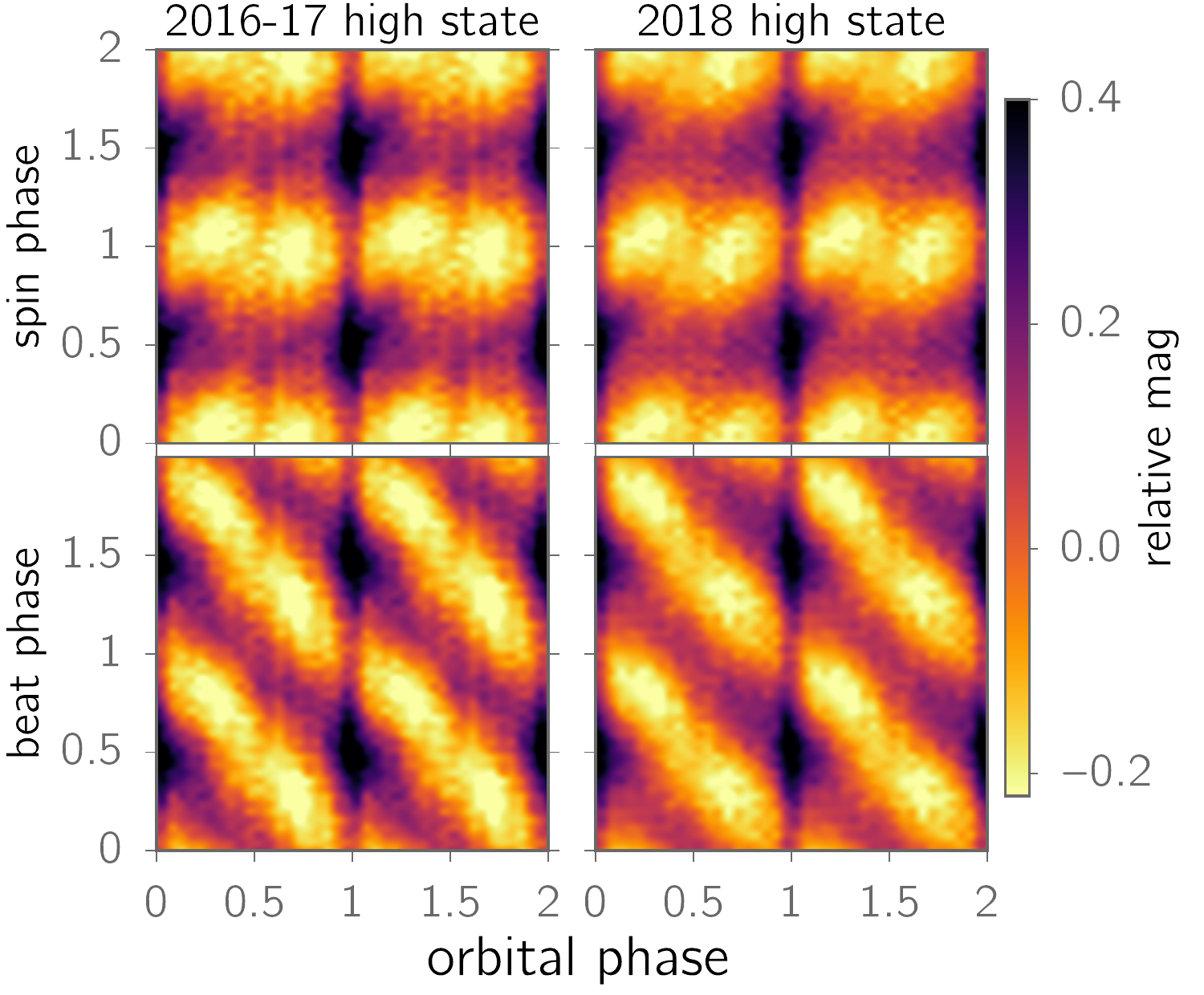}
    \caption{Two-dimensional, stroboscopic light curves of each low state (top row) and high state (bottom row) in our dataset; note that the dynamic range represented by the colormap is different for the two rows. A horizontal slice through one of the panels gives the orbital light curve at a fixed spin or beat phase, while a vertical slice yields the spin or beat light curve at a fixed orbital phase.  During the low states, the eclipses were deepest when the beat phase was $\sim$0.6.\label{fig:stroboscopic}}
\end{figure}

We show stroboscopic light curves of both the spin and beat modulations in Fig.~\ref{fig:stroboscopic}. We define beat phase 0.0 as the accretion geometry during which the upper magnetic pole tilts away from the companion.\footnote{We presume that this occurs when the spin pulse occurs at an orbital phase of 0.0.} The complex structure of these stroboscopic light curves yields insight into the interplay between FO Aqr's spin, beat, and orbital periods. For example, in the high state, the eclipse depth varies dramatically as a function of spin phase, with maximum depth occurring at $\phi_{spin} \sim 0.5$, the minimum of the spin pulse. This is fully consistent with the eclipse depth being diluted by light from the spin pulse, in agreement with the widely accepted viewing geometry in which the upper accretion curtain produces the spin pulse and is never eclipsed by the companion. Likewise, the stroboscopic, low-state beat light curves show two peaks near $\phi_{beat} \sim 0.3$ and $\phi_{beat} \sim 0.8$, as might be expected from a magnetic dipole accreting from a stationary source in the binary rest frame.

The stroboscopic light curves in Fig.~\ref{fig:stroboscopic} also offer an explanation for why the high-state \textit{K2} orbital waveform in Fig.~\ref{2016eclipses} is double-humped: the amplitude of the spin pulse changes across the orbit during the high state. Although phase-averaging the light curve to the orbit will smear out individual pulses, this technique does not compensate for the increased contribution from the spin pulses near orbital phases $\sim$0.3 and $\sim$0.7, creating the appearance of an orbital brightening. If the system could be observed at the minimum of the spin pulse across the entire orbit, the out-of-eclipse light curve would not show the double-humped waveform (Fig.~\ref{fig:stroboscopic}).

\end{document}